\documentclass[superscriptaddress,amsmath,amssymb,jcp,aip,]{revtex4-1}

\usepackage{graphicx,dcolumn,bm,xcolor,microtype,multirow,amscd,amsmath,amssymb,amsfonts,physics,longtable,wrapfig,xspace}
\usepackage[version=4]{mhchem}

\usepackage[utf8]{inputenc}
\usepackage[T1]{fontenc}
\usepackage{txfonts}
\usepackage{grffile}

\usepackage[normalem]{ulem}

\definecolor{darkgreen}{RGB}{0, 180, 0}

\usepackage[
	colorlinks=true,
    citecolor=blue,
    breaklinks=true
	]{hyperref}
\newcommand{\ePT}[1]{e_{#1}}

\newcommand{\Nelec}{N_\text{e}}
\newcommand{\Nbas}{N_\text{bas}}
\newcommand{\Nnuc}{N_\text{n}}

\newcommand\QP{\textsc{quantum package}\xspace}

\newcommand{\EDMC}{E_\text{DMC}}
\newcommand{\ESDDMC}{E_\text{SD-DMC}}
\newcommand{\EMDDMC}{E_\text{MD-DMC}}
\newcommand{\EHFKS}{E_\text{HF}^\text{B3LYP}}
\newcommand{\Efit}{E_\text{fit}}

\newcommand{\EPT}{E_\text{PT2}}
\newcommand{\Evar}{E_\text{var}}

\newcommand{\cMO}[1]{C_{#1}}

\newcommand{\AO}[1]{\Tilde{\chi}_{#1}}
\newcommand{\pAO}[1]{\chi_{#1}}

\newcommand{\pMO}[1]{\phi_{#1}}

\newcommand{\hH}{\Hat{H}}
\newcommand{\hO}{\Hat{O}}
\newcommand{\hT}{\Hat{T}}
\newcommand{\hVee}{\Hat{V}_\text{ee}}
\newcommand{\hVne}{\Hat{V}_\text{ne}}
\newcommand{\hVnn}{\Hat{V}_\text{nn}}

\newcommand{\kalpha}{{\ket{D_\alpha}}}
\newcommand{\kD}{\ket*{D}}
\newcommand{\kI}{\ket{D_I}}

\newcommand{\br}{{\mathbf{r}}}
\newcommand{\bt}{{\mathbf{t}}}
\newcommand{\bk}{{\mathbf{k}}}
\newcommand{\bg}{{\mathbf{g}}}
\newcommand{\bR}{{\mathbf{R}}}
\newcommand{\bT}{{\mathbf{T}}}
\newcommand{\bO}{{\mathbf{0}}}
\newcommand{\bG}{{\mathbf{G}}}
\newcommand{\bK}{{\mathbf{K}}}
\newcommand{\bA}{{\mathbf{A}}}
\newcommand{\bB}{{\mathbf{B}}}
\newcommand{\bC}{{\mathbf{C}}}
\newcommand{\ba}{{\mathbf{a}}}
\newcommand{\bb}{{\mathbf{b}}}
\newcommand{\bc}{{\mathbf{c}}}

\newcommand{\tabc}[1]{\multicolumn{1}{c}{#1}}

\newcommand{\fnm}{\footnotemark}
\newcommand{\fnt}{\footnotetext}

\newcommand{\SI}{\textcolor{blue}{supplementary material}}
\newcommand{\ie}{\textit{i.e.}}
\newcommand{\eg}{\textit{e.g.}}
\newcommand{\cell}[3]{{#1} \times {#2} \times {#3}}

\newcommand{\ii}{\mathrm{i}}

\begin{document}	
\title{Towards a Systematic Improvement of the Fixed-Node Approximation in Diffusion Monte Carlo for Solids{ - A Case Study in Diamond}}

\author{Anouar Benali}
\email{benali@anl.gov}
\affiliation{Computational Sciences Division, Argonne National Laboratory, Argonne, IL 60439, United States}
\author{Kevin Gasperich}
\affiliation{Department of Chemistry, University of Pittsburgh, Pittsburgh, Pennsylvania 15260, United States}
\author{Kenneth D.~Jordan}
\affiliation{Department of Chemistry, University of Pittsburgh, Pittsburgh, Pennsylvania 15260, United States}
\author{Thomas Applencourt}
\affiliation{Argonne Leadership Computing Facility, Argonne National Laboratory, Argonne, IL 60439, United States}
\author{Ye Luo}
\affiliation{Computational Sciences Division, Argonne National Laboratory, Argonne, IL 60439, United States}
\author{M.~Chandler Bennett}
\affiliation{Materials Science and Technology Division, Oak Ridge National Laboratory, Oak Ridge, TN 37831, United States\footnote{This manuscript has been authored by UT-Battelle, LLC under Contract No. DE-AC05-00OR22725 with the U.S. Department of
Energy. The United States Government retains and the publisher, by accepting the article for publication, acknowledges that the
United States Government retains a non-exclusive, paid-up, irrevocable, worldwide license to publish or reproduce the published
form of this manuscript, or allow others to do so, for United States Government purposes. The Department of Energy will provide
public access to these results of federally sponsored research in accordance with the DOE Public Access Plan
(http://energy.gov/downloads/doe-public-access-plan).}}
\author{Jaron T.~Krogel}
\affiliation{Materials Science and Technology Division, Oak Ridge National Laboratory, Oak Ridge, TN 37831, United States}
\author{Luke Shulenburger}
\affiliation{HEDP Theory Department, Sandia National Laboratories, Albuquerque, New Mexico 87185, USA}
\author{Paul R.~C.~Kent}
\affiliation{Center for Nanophase Materials Sciences, Oak Ridge National Laboratory, Oak Ridge, TN 37831, United States}
\affiliation{Computational Sciences and Engineering Division, Oak Ridge National Laboratory, Oak Ridge, TN 37831, United States}
\author{Pierre-Fran\c{c}ois Loos}
\affiliation{Laboratoire de Chimie et Physique Quantiques, Universit\'e de Toulouse, CNRS, UPS, {Toulouse,} France}
\author{Anthony Scemama}
\affiliation{Laboratoire de Chimie et Physique Quantiques, Universit\'e de Toulouse, CNRS, UPS, {Toulouse,} France}
\author{Michel Caffarel}
\email{caffarel@irsamc.ups-tlse.fr}
\affiliation{Laboratoire de Chimie et Physique Quantiques, Universit\'e de Toulouse, CNRS, UPS, {Toulouse,} France}

\begin{abstract}
While Diffusion Monte Carlo (DMC) is in principle an exact stochastic method for \textit{ab initio} electronic structure calculations, in practice the fermionic sign problem necessitates the use of the fixed-node approximation and trial wavefunctions with approximate nodes (or zeros) must be used.
This approximation introduces a variational error in the energy that potentially can be tested and systematically improved. Here,
we present a computational method that produces trial wavefunctions with systematically improvable nodes for DMC calculations of
periodic solids. These trial wavefunctions are efficiently generated with the configuration interaction using a perturbative
selection made iteratively {(CIPSI)} method.
{A simple protocol in which both exact and approximate results for finite supercells are used 
to extrapolate to the thermodynamic limit is introduced.}
This approach is illustrated on the case of the carbon diamond using Slater-Jastrow trial wavefunctions 
including up to one million Slater determinants. {Fixed-node DMC energies obtained with such large expansions are much improved and
the fixed-node error is found to decrease monotonically and smoothly as a function of the number of determinants in the trial wavefunction, a property
opening the way to a better control of this error. The cohesive energy extrapolated to the thermodynamic limit is in close agreement 
with the estimated experimental value. Interestingly, this is also the case at the single-determinant level,
thus indicating a very good error cancelation in carbon diamond
between the bulk and atomic total fixed-node energies when using single-determinant nodes.}
\end{abstract}

\maketitle

\section{Introduction}

A faithful and quantitative first principles (\ie, \textit{ab initio}) description of the electronic structure of solids is one of
the greatest challenges of material science. For weakly correlated materials (most simple metals, semiconductors, and insulators),
modern approximations to exact density-functional theory (DFT), \cite{Hohenberg_1964,Kohn_1965,ParrBook} and approximate many-body
perturbation theory (MBPT) approaches (such as the $GW$ approximation \cite{Hedin_1965,Strinati_1980} and the Bethe-Salpeter
equation \cite{Salpeter_1951,Strinati_1984,Strinati_1988}) generally yield ground- and excited-state properties in good agreement
with experiment.~\cite{Martin_2004,Martin_2016} In sharp contrast, for strongly correlated systems (\eg, transition metal oxides)
for which electrons can no longer be treated as weakly interacting quasiparticles, the approaches mentioned above may dramatically
fail, even qualitatively.\cite{DiSabatino_2016,DiSabatino_2019}  
{An alternative path to describe correlation effects in solids is to resort to wave-function-based correlated methods, as done in a number of recent works [e.g. {second order Møller-Plesset Perturbation theory (MP2),\cite{Pisani_2005} Coupled Cluster Single and Double (CCSD) with perturbative Triple (CCSD(T)), and the Equation Of Motion-CCSD (EOM-CCSD)\cite{McClain_2017,Gruber_2018}}]. However, as single-reference approaches they are of limited use for strongly correlated materials.}  Designing efficient and
general computational approaches with controlled and testable approximations is therefore of great interest.

Quantum Monte Carlo (QMC) methods have been developed to treat weakly through strongly correlated systems while invoking
few approximations. These include variational Monte Carlo (VMC) and diffusion Monte Carlo (DMC) in continuous space
\cite{Foulkes_2001,Austin_2012},  and full configuration interaction QMC (FCIQMC) \cite{Booth_2009,Booth_2013}
and auxiliary field QMC (AFQMC)\cite{Zhang_1997,Zhang_2003} in determinant space. These methods utilize stochastic integration to
help treat the complexity of the many-body electronic structure problem, and are applicable to both solids and isolated molecules.
Importantly it is becoming possible to compare these and other many-body methods with each other on equivalent problems and to
both validate their results and focus improvements.\cite{Motta_2017,Kiel_2020} In this study we will focus on DMC, which is one of the
most widely used QMC approach for calculating the electronic structure of solids. \cite{Foulkes_2001,Lester_2009} DMC enjoys a
relatively low computational scaling with system size and is well adapted to massive parallelism.
\cite{Nakano_2020,Scemama_2013b,Needs_2020,Kim_2018,Kent_2020} 
For bosons, DMC can deliver the exact energy to an arbitrarily small error. However, for fermionic problems, {because of the uncontrolled fluctuations of the wavefunction sign
a DMC implementation that does not address the sign problem is not stable in time}. \cite{Reynolds_1982,Ceperley_1991} 
To address the fermion sign problem and correctly obtain
the required fermionic symmetry, practical DMC calculations must rely on the fixed-node approximation: the fermionic constraints
are fulfilled by imposing the zeros (nodes) of the wavefunction to be those of an approximate antisymmetric (fermionic) trial
function. However, because these nodes are approximate, the DMC energy has a residual variational error, referred to as the
fixed-node error. Devising an efficient strategy to improve the trial wavefunction nodes \textit{in a systematic way} and, thus, {better controlling the only source of error of the method, brings us closer to achieving a genuine general from-first-principles theory, which is one of the major challenges of DMC.}

One advantage of real space QMC methods is their ability to flexibly use different forms of trial {wavefunctions}. Since only the
values and derivatives of the wavefunction are required there is no need in principle to choose forms that are suited for
conventional numerical integration. While it is now standard to utilize sophisticated multideterminant trial wavefunctions in
molecules, most solid state applications of DMC use a single {Slater} determinant (SD) trial wavefunction. A Jastrow factor is
used to incorporate dynamical correlations and improve the wavefunction overall, but it does not change the nodes. Although SD-DMC
calculations have proven valuable in addressing a wide range of problems in chemistry and materials science, for example obtaining
near chemical accuracy in van der Waals binding\cite{Benali_2014,Dubecky_2016}, this accuracy is far from general. A potentially general
strategy to address this problem, which has been successfully applied to atoms and molecules, is to employ multideterminant (MD)
trial functions. \cite{Giner_2013,Morales_2012,Petruzielo_2012,Caffarel_2016b}. However this requires an efficient method to
generate the multideterminant expansion.

In this work, we present a scheme to perform multideterminant DMC calculations in solids using configuration interaction (CI) based
trial functions. We employ a selected CI (SCI) approach in which only the most important Slater determinants from the full CI
(FCI) space are identified and selected. 
{Although SCI does not suppress the exponential wall of conventional CI calculations, it may { be deferred sufficiently}
to allow practical calculations.}
In the case of atomic and molecular systems, several
successful applications of SCI trial functions in DMC calculations have been reported.
\cite{Scemama_2019,Scemama_2018a,Scemama_2018b,Caffarel_2016a,Caffarel_2016b,Scemama_2014,Giner_2013,Dash_2018,Dash_2019} Here, we
use the configuration interaction using a perturbative selection made iteratively (CIPSI) algorithm, \cite{Huron_1973} as
implemented in the {\QP} code \cite{Garniron_2019b} to generate the trial functions for DMC calculations on periodic systems. This approach
is illustrated by computing the cohesive energy of carbon diamond in the thermodynamic limit. We use the QMCPACK quantum Monte
Carlo package \cite{Kim_2018,Kent_2020} for the DMC calculations. Although diamond is not a particularly challenging system
with respect to the role of electron correlation, it is a valuable test system since its cohesive energy is accurately known and
pseudopotential accuracy is not a major concern.
{This work may thus be considered as a first step toward the study of more difficult solids where greater correlation effects and the generation of CIPSI expansions may be more challenging.}

The paper is organized as follows. In Sec.~\ref{sec:theory}, the theoretical aspects of the approach are presented. A brief
summary of the CIPSI method is given and its extension to the case of periodic systems is described in depth. The computational
details are provided in Sec.~\ref{sec:compdet}. Results for diamond, both at the single- and multideterminant levels, are
reported and discussed in Sec.~\ref{sec:results}. Finally, some concluding remarks are given in Sec.~\ref{sec:conclusion},
including areas where further improvements are desired. Unless otherwise stated, atomic units are used throughout this paper.

\section{Theory}\label{sec:theory}

\subsection{The CIPSI algorithm}\label{sec:CIPSI}

In the present manuscript, the CIPSI algorithm developed for finite-size atomic or molecular systems is generalized to the case of
periodic solids. CIPSI is one of the variants of the broad class of methods known as SCI. Selecting determinants in the CI space
is a natural idea, and many SCI variants have been developed under various acronyms and implementations over the last five
decades.
\cite{Bender_1969,Whitten_1969,Huron_1973,Buenker_1974,Buenker_1975,Buenker_1978,Evangelisti_1983,Cimiraglia_1985,Cimiraglia_1987,Illas_1988,Povill_1992,Abrams_2005,Bunge_2006,Bytautas_2009,Giner_2013,Caffarel_2014,Giner_2015,Garniron_2017b,Caffarel_2016a,Caffarel_2016b,Holmes_2016,Sharma_2017,Holmes_2017,Chien_2018,Scemama_2018a,Scemama_2018b,Loos_2018b,Garniron_2018,Evangelista_2014,Schriber_2016,Schriber_2017,Liu_2016,Per_2017,Ohtsuka_2017,Zimmerman_2017,Li_2018,Ohtsuka_2017,
Coe_2018,Loos_2019,Loos_2020a,Loos_2020b} SCI methods are ordinary CI approaches except that determinants are not chosen based on
predetermined occupation or excitation criteria, but are instead selected step by step based on their estimated contribution to
the FCI energy and/or wavefunction. It is a particularly successful approach since it is well-recognized that, in a predefined
subspace of determinants {for example, single and double excitations with respect to the Hartree-Fock (HF)  reference}, only a small fraction of
them make a non-negligible contribution to the wavefunction. \cite{Bytautas_2009,Anderson_2018} {Therefore, an \emph{on-the-fly} selection of determinants has been proposed in the late 1960s by Bender and Davidson,\cite{Bender_1969} as well as by Whitten and
Hackmeyer.\cite{Whitten_1969}} The main advantage of SCI methods is that no \emph{a priori} assumption is made on the
type of determinants needed to describe the electronic correlation effects. Therefore, at the potentially greater price of a blind
and automated calculation, {a} SCI calculation is less biased by the user's understanding of the problem's complexity (for example,
when choosing the active space orbitals for a particular problem).

The CIPSI approach was developed in 1973 by Huron, Rancurel, and Malrieu. \cite{Huron_1973} In recent years, two of us have
revived this approach \cite{Giner_2013} and developed a very efficient and massively parallel version of the algorithm which has
been implemented in {\QP}. \cite{Garniron_2019b} Briefly, at each iteration $n$, CIPSI selects some external determinants $\kalpha$
not present in the current zeroth-order wavefunction
\begin{equation}
    \ket*{\Psi_0^{(n)}} = \sum_{I} c_I^{(n)} \kI,
\end{equation}
where $\kI$ is an internal determinant. Starting, for example, with the HF determinant at $n=0$, the external determinants are
selected using a criterion based on the estimated gain in correlation energy evaluated using second-order perturbation theory that
would result from the inclusion of $\kalpha$. Denoting the second-order correction as
\begin{equation}
    \ePT{\alpha}^{(n)} = \frac{ \abs*{\mel*{\Psi_0^{(n)}}{\hH}{D_\alpha}}^2 }{\Evar^{(n)} - \mel*{D_\alpha}{\hH}{D_\alpha}},
\end{equation}
where
\begin{equation}
    \Evar^{(n)} = \frac{
    \mel*{\Psi_0^{(n)}}{\hH}{\Psi_0^{(n)}}
    }{
    \braket*{\Psi_0^{(n)}}{\Psi_0^{(n)}}
    }
\end{equation}
 is the variational energy of the wavefunction at the $n$th iteration, a number of external determinants associated with the
greatest $\ePT{\alpha}^{(n)}$ are incorporated into the variational space and the Hamiltonian is diagonalized to give
$\ket*{\Psi_0^{(n+1)}}$ and $\Evar^{(n+1)}$. In practice, the size of the variational wavefunction is roughly doubled at each
iteration. Next, the second-order Epstein-Nesbet energy correction to the variational energy (denoted as $\EPT^{(n)}$) is computed
by summing up the contributions of all external determinants
\begin{equation}
    \EPT^{(n)} = \sum_{\alpha} \ePT{\alpha}^{(n)},
\label{PT2}
\end{equation}
and the total CIPSI energy is given by
\begin{equation}
    E_\text{CIPSI}^{(n)} = \Evar^{(n)} + \EPT^{(n)}.
\end{equation}
The algorithm is then iterated until some convergence criterion (for example, $|\EPT^{(n)}| \le \epsilon$) is met. For simplicity,
in the following the superscript denoting the largest iteration number attained will be dropped from the various expressions. When
the number of determinants in the variational space gets large, computing the second-order correction $\EPT$ by adding up all
contributions $\ePT{\alpha}$ to the sum becomes computationally unfeasible. To perform this formidable task, a simple and efficient hybrid
stochastic-deterministic algorithm has been developed. \cite{Garniron_2017b} In short, the leading contributions to the sum in
Eq.~\eqref{PT2} are exactly computed (deterministic part) while {the} very large number of small residual contributions are sampled
via a Monte Carlo algorithm (stochastic part).

\subsection{CIPSI for periodic systems}

We now present the extension of CIPSI to periodic solids. As in any CI calculation, we must define (i) the system (number of
electrons, charge and positions of the nuclei), (ii) the Hamiltonian, and (iii) the one-electron basis set. The calculation of the
one-and two-electron integrals required for the computation of the Hamiltonian matrix elements then needs to be discussed.

\subsubsection{Supercells}

In practice, an infinite solid is modeled by a finite supercell obtained by replicating a given primitive cell $n_i$ times in each
Cartesian direction (\ie, $i=x,y,z$). Accordingly, we will label a supercell as $\cell{n_x}{n_y}{n_z}$. In the present study, we
will restrict ourselves to the simplest case of a cubic primitive cell of side $L$ with arbitrary $n_x$, $n_y$, and $n_z$ values.
The supercell is then a rectangular cuboid of volume $\Omega = N \times L^3$, where $N = n_x n_y n_z$ is the total number of
primitive cells replicated to build the supercell. 

The supercell being defined, we are led back to an ordinary CI calculation consisting of the set of electrons and nuclei present
in the supercell and subject to an external periodic electric potential. We denote $\bR_I$ and $Z_I$ the position and charge of
the of $I$th nuclei of the primitive cell ($I=1,\ldots,\Nnuc$). The actual system is then composed of $N \times \Nnuc$ nuclei at
positions $\bR_I + t_a \ba + t_b \bb + t_c \bc$, and of $N \times \Nelec$ electrons, where $\Nelec$ is the number of electrons of
the primitive cell. Here, $\bt = t_a \ba + t_b \bb + t_c \bc = (t_a,t_b,t_c)$ is the lattice translation vector, and the triplet
of integers $(t_a,t_b,t_c)$ takes all the values needed to generate the supercell through translations of the primitive cell along
its unit vectors $(\ba,\bb,\bc)$. Similarly, the supercell translation vector is defined as $\bT = T_A \bA + T_B \bB + T_C \bC =
(T_A,T_B,T_C)$, where, in the case of a cubic primitive cell, $(\bA,\bB,\bC) = (L\ba,L\bb,L\bc)$ are the corresponding unit
translation vectors of the supercell.

We emphasize that, in contrast to effective one-body theories such as HF or DFT, many-body
electronic structure calculations explicitly taking into account the electron-electron interaction,
such as the ones performed here, cannot be restricted to the primitive cell if accurate properties are to be obtained.\cite{Foulkes_2001}
In one-body theories the thermodynamic limit can be reached by indefinitely improving the ${\bk}$-point sampling of the first
Brillouin zone of the primitive cell. In presence of electron-electron interaction the
translation of \textit{individual} electrons is no longer a symmetry of the Hamiltonian (${\bk}$ is no longer a good quantum number)
and the use of supercells is mandatory (see, e.g., Refs \onlinecite{Foulkes_2001} and \onlinecite{Kolorenvc_2011} for a discussion of this aspect).

\subsubsection{Hamiltonian}

The supercell electronic Hamiltonian
\begin{equation}
	\hH_N = \hT + \hVne + \hVee + \hVnn
\end{equation} 
has a standard form, except that the electron-electron, electron-nucleus, and nucleus-nucleus Coulombic potentials, $\hVee$,
$\hVne$, and $\hVnn$, respectively, are now periodized to model the interaction of the electrons and nuclei belonging to the
supercell with the infinite set of replicas associated with their periodic images, \ie,
\begin{equation}
\label{eq:H1}
\begin{split}
	\hH_N 
	& = -\frac{1}{2} \sum_{i} \nabla_i^2 
	- \sum_{\bT} \sum_{i} \sum_{J} 
	\frac{Z_J}{ \abs*{\br_i - \bR_J + \bT } }
	\\
	& + \frac{1}{2} \sum_{\bT} \sideset{}{'}\sum_{ij}
	\frac{1}{ \abs*{\br_i - \br_j + \bT } }
	+ \frac{1}{2} \sum_{\bT} \sideset{}{'}\sum_{IJ}
	\frac{Z_I Z_J}{ \abs*{ \bR_I - \bR_J + \bT }},
\end{split}
\end{equation}
where $\br_i$ is the position of the $i$th electron, $\sum_{\bT} \equiv \sum_{T_A = - \infty}^{\infty} \sum_{T_B = -
\infty}^{\infty} \sum_{T_C = - \infty}^{\infty}$, and the prime symbol indicates that the self-interaction term $i=j$ or $I = J$
has to be excluded when $\bT = \bO = (0,0,0)$.

As is well known, periodic Coulomb potentials are mathematically ill-defined due to the long-range character of the Coulomb
interaction. The periodic infinite sums in Eq.~\eqref{eq:H1} not only converge very slowly but they are also conditionally
convergent, meaning that the result depends on the order of summation. A specific and careful mathematical treatment has to be
introduced in order to provide a meaningful answer. The standard solution, and the one we employ here for the three Coulomb
potentials in Eq.~\eqref{eq:H1}, is to resort to the Ewald summation technique. 

Applying the Ewald summation technique, for example, to the nucleus-nucleus term $\hVnn$ in Eq.~\eqref{eq:H1} enables one to
compute this term as a sum of two contributions: a  short-range contribution in real space and a short-range in reciprocal space.
Both contributions are expressed as rapidly converging infinite sums, thus leading to a very fast and efficient calculation of the
potential. Explicitly, it is given by
\begin{equation}
\label{eq:H2}
\begin{split}
	\hVnn 
	& = \frac{1}{2} \sum_{\bT} \sideset{}{'}\sum_{IJ} \frac{Z_I Z_J}{ \abs*{\bR_I - \bR_J + \bT } } 
	\text{erfc} \qty[ \frac{\abs*{\bR_I - \bR_J + \bT }} {\sqrt{2} \sigma} ]
	\\
	& + \frac{2\pi}{\Omega} \sum_{\bG \ne \bO} \frac{e^{-\frac{\sigma^2 \bG^2}{2}}} {\bG^2} \abs{\sum_I Z_I e^{\ii \bG \cdot \bR_I}}^2 
	- \frac{ \sum_{I} Z_I^2}{\sqrt{2 \pi} \sigma} - \frac{\pi \sigma^2}{\Omega} \qty(\sum_{I} Z_I \bR_I)^2,
\end{split}
\end{equation}
where $\text{erfc}(x)$ is the complementary error function, $\bG$ refers to the set of quantized reciprocal vectors of the
supercell defined by the condition $e^{\ii \bG \cdot \bT} = 1$, and $\sigma$ is a small positive parameter chosen to facilitate
convergence. Additional details can be found, for example, in Refs.~\onlinecite{deLeeuw_1980,AllenBook}.

\subsubsection{Basis functions}
In this work, the one-electron basis functions are chosen to be crystalline Gaussian-based atomic orbitals 
\begin{equation}
\label{eq:AO}
	\pAO{\mu\bk}(\br) = \sum_{\bT} e^{\ii \bk \cdot \bT} \AO{\mu}(\br + \bT),
\end{equation}
\ie, the periodized (or translationally-symmetry-adapted) version of the (localized) Gaussian atomic orbitals $\AO{\mu}(\br)$ from
the supercell. In Eq.~\eqref{eq:AO}, the crystal momentum vector $\bk$ is chosen within the first Brillouin zone of the primitive
cell, and it is sampled from a Monkhorst-Pack grid, \cite{Monkhorst_1976} an evenly-spaced rectangular grid in reciprocal space
[see Sec.~\eqref{sec:PBC}]. The atomic index $\mu$ is referred to as the band index after periodization.

The molecular orbitals of the system are then defined as
\begin{equation}
\label{eq:MO}
	\pMO{p\bk}(\br) = \sum_{\mu=1}^{\Nbas} \cMO{\mu p}(\bk) \, \pAO{\mu\bk}(\br),
\end{equation}
where $\Nbas$ is the number of basis functions, and the molecular orbital coefficients $\cMO{\mu p}(\bk)$ are now
momentum-dependent due to the translational-symmetry adaptation of the basis functions. Once again, we emphasize that, because of
the explicit treatment of the electron-electron interaction, $\bk$ is no longer a good quantum number and the orbitals defined
above do not have the correct translational symmetry of the problem. However, choosing such a representation is particularly
convenient in practice since it allows to take full advantage of the techniques and codes developed for the effective one-particle
theories, such as HF and DFT.

\subsubsection{One- and two-electron integrals}
The Hamiltonian and the one-electron basis set having been defined, the next step in a CI calculation is to evaluate the
Hamiltonian matrix elements between Slater determinants. This requires evaluation of one- and two-electron integrals. It is easy
to see from the expressions of the Hamiltonian [see Eqs.~\eqref{eq:H1} and \eqref{eq:H2}] and basis functions [see
Eq.~\eqref{eq:AO}] that this can be accomplished by calculation of the integrals over Fourier transforms of the product of Gaussian
functions for which fast and reliable algorithms have been developed. \cite{Gill_1989,Gill_1991,McClain_2017}

The only new aspect with respect to standard CI implementations is the use of complex-valued orbitals, integrals and Hamiltonian
matrix elements. Let us denote the two-electron integrals as
\begin{equation}
\braket*{p\bk_p q\bk_q}{r\bk_r s\bk_s} = \iint_{\Omega^2} d\br_1 d\br_2 \pMO{p\bk_p}^*(\br_1)\pMO{q\bk_q}^*(\br_2) \frac{1}{r_{12}} \pMO{r\bk_r}(\br_1) \pMO{s\bk_s}(\br_2).
\end{equation}
Since the one-particle basis functions are invariant with respect to the primitive lattice translation vectors $\bt$, the
two-electron integrals must conserve crystal momentum, \ie, $\bk_p + \bk_q = \bk_r + \bk_s + \bg$, where $\bg$ is a reciprocal
lattice vector of the primitive cell.

When real-valued orbitals are used, a given two-electron integral is symmetric with respect to permutations of the orbital indices
$p$ and $r$, or $q$ and $s$, as well as with the permutation of the electron labels 1 and 2, thus resulting in a 8-fold symmetry
of the set of two-electron integrals. Consequently, the $\Nbas^4$ four-index integrals can be mapped to a set of approximately
$\Nbas^4/8$ unique integral values. \cite{SzaboBook} However, when one considers complex-valued orbitals, the integral
$\braket*{p\bk_p q\bk_q}{r\bk_r s\bk_s}$ is no longer invariant with the permutations of $p$ and $r$, or $q$ and $s$, so the
number of unique integral values scales as $\Nbas^4/2$ (exchanging the electron indices 1 and 2 still leaves the value of the
integral unchanged). When storing these complex-valued integrals, it is useful to recognize that exchanging $p$ with $r$, or $q$
with $s$ changes the integral value in a non-trivial way. However, exchanging both of these pairs simultaneously yields the
complex conjugate of the original value, \ie, $\braket*{p\bk_p q\bk_q}{r\bk_r s\bk_s}^* = \braket*{r\bk_r s\bk_s}{p\bk_p q\bk_q}$.
This allows one to save an additional factor of two in the storage requirements; one then only needs to store $\Nbas^4/4$
complex-valued two-electron integrals.

\subsubsection{Hamiltonian matrix elements}

In order to implement FCI or any of its lower-cost alternatives, one must be able to evaluate matrix elements of the Hamiltonian
in the space of Slater determinants. If the determinants are all built from the same set of orthonormal spin-orbitals (which is
the case here), then one can evaluate matrix elements via the usual Slater-Condon rules (see, \eg, Ref.~\onlinecite{SzaboBook} and
Ref.~\onlinecite{Scemama_2013a} for an efficient implementation in a determinant-driven context).

\subsection{Boundary conditions and twist averaging}\label{sec:PBC}
In order to reduce finite-size effects and, to accelerate the convergence to the thermodynamic limit (\ie, $N \to \infty$), it is
advantageous to exploit the freedom in the type of periodic boundary conditions of the supercell. By judiciously choosing the
electron momenta $\bk$ of the basis functions [see Eq.~\eqref{eq:AO}], the boundary conditions can be easily implemented.
Translating one electron of the supercell by a superlattice vector (say, $\bA = L \ba)$ generates a phase factor, $e^{i L \bk
\cdot \ba}$, for each of the orbitals of the CI determinants. Accordingly, a global phase factor common to all determinants is
obtained whenever all these individual phase factors are made equal, that is
\begin{equation}
	e^{\ii L \bk_1 \cdot \ba} = e^{\ii L \bk_2 \cdot \ba} = \ldots = e^{\ii \theta},
\end{equation}
where $\theta$ is some arbitrary angle (or twist) between $-\pi$ and $\pi$. These conditions are fulfilled when momenta are chosen
uniformly spaced in the first Brillouin zone of the lattice (that is, by using a Monkhorst-Pack grid \cite{Monkhorst_1976}) and
shifted by a common vector $\bK$
\begin{subequations}
\begin{align}
	\label{eq:mpa1}
	\bk_{i_a} & = \bK + \frac{i_a}{t_a} \frac{ \bb \times \bc }{ \ba \cdot (\bb \times \bc)},
	\\
	\label{eq:mpa2}
	\bk_{i_b} & = \bK + \frac{i_b}{t_b} \frac{ \bc \times \ba }{ \bb \cdot (\bc \times \ba)},
	\\
	\label{eq:mpa3}
	\bk_{i_c} & = \bK + \frac{i_c}{t_c} \frac{ \ba \times \bb }{ \bc \cdot (\ba \times \bb)},
\end{align}
\end{subequations}
with $i_l = 0,\ldots,t_l$ ($l=a,b,c$) is the twist index. In this case, we have $\kD \rightarrow e^{\ii \theta} \kD$, with
$\theta= \bK \cdot \ba$. Boundary conditions can be varied with $\theta$ from $-\pi$ to $+\pi$, $\theta = 0$ and $\theta = \pm\pi$
corresponding to periodic and anti-periodic boundary conditions, respectively.

For a given system size, a property $O$ can be computed by averaging out its values for different $\bK$ values (or twists
$\theta$). In the limit of an infinite set of sampling values, we have 
\begin{equation}
	\expval{O} = \frac{1}{2\pi} \int_{-\pi}^{\pi} d\theta \mel{\Psi_{\theta}}{\hO}{\Psi_{\theta}},
\end{equation}
where $\Psi_{\theta}$ is the exact wavefunction of the system for the corresponding boundary condition. In practice, the integral
is computed as a finite sum over shifted Monkhorst-Pack grids, as expressed by Eqs.~\eqref{eq:mpa1}, \eqref{eq:mpa2}, and
\eqref{eq:mpa3}. The main effect of twist-averaging is to suppress the major part of the one-body shell effects in the filling of
single-particle states. Note that each value of $\theta$ requires an independent calculation, the total computational cost is then
proportional to the number of twists.

\section{Computational details}\label{sec:compdet}
The DMC method is employed to compute ground-state energies within the fixed-node approximation. The application of DMC to solids
being now well documented, we refer the interested reader to the existing literature for the theoretical background (see, for
example, Refs.~\onlinecite{Foulkes_2001,Kolorenvc_2011,Kim_2018}). Calculations were performed using QMCPACK.\cite{Kim_2018,Kent_2020} The integrals over periodized Gaussian-type orbitals are carried out using the PySCF
program, \cite{Sun_2017} which has been interfaced with {\QP}. \cite{Garniron_2019b} {Plane wave trial wavefunctions were generated with the QUANTUM ESPRESSO package\cite{giannozzi09}}. 

To model carbon diamond, we used a primitive cell with two carbon atoms separated by $1.545$ {\AA} (approximately, $3.568$ {\AA}
for the lattice constant). In order to compute the cohesive energy in the thermodynamic limit and to correct energies for
finite-size effects, we consider supercells made of 16 ($\cell{2}{2}{2}$), 54 ($\cell{3}{3}{3}$) and 128 ($\cell{4}{4}{4}$)
atoms with converged twist-averaged (TA) boundary conditions with a total of $216$, $64$ and $27$ twists respectively.
\cite{Lin_2001} All calculations are carried out using the Burkatzki-Filippi-Dolg (BFD) \cite{Burkatzki_2007,Burkatzki_2008}
effective core potentials and the associated $2s2p1d$ or $3s3p2d1f$ contracted Gaussian-type orbital (GTO) basis sets (BFD-vDZ or
BFD-vTZ, respectively). {For the plane wave trial wavefunction, consistently with Ref.~\onlinecite{Shin_2014}, a 200Ry cut-off was applied to the kinetic energy.}   Both SD-DMC and MD-DMC calculations used a time step of $0.001$ a.u. and Casula's T-moves for
pseudopotential {evaluation}.
\cite{Casula_2006} 

The trial wavefunctions for DMC consist either of a single determinant or multideterminant expansion determined using CIPSI multiplied
by a Jastrow factor with one, two, and three body terms. The parameters for the one- and two-body terms were represented by
B-splines and the three-body term by a polynomial. These parameters were optimized in VMC through a variant of the linear method
of Umrigar \textit{et al.} \cite{Umrigar_2007} Note that the coefficients of the CIPSI expansion were not optimized in VMC and
therefore in this study the nodal surface is wholly determined by the CIPSI expansion.

\section{Results}\label{sec:results}

\subsection{Single-determinant fixed-node DMC}

\begin{figure}
\includegraphics[width=\linewidth]{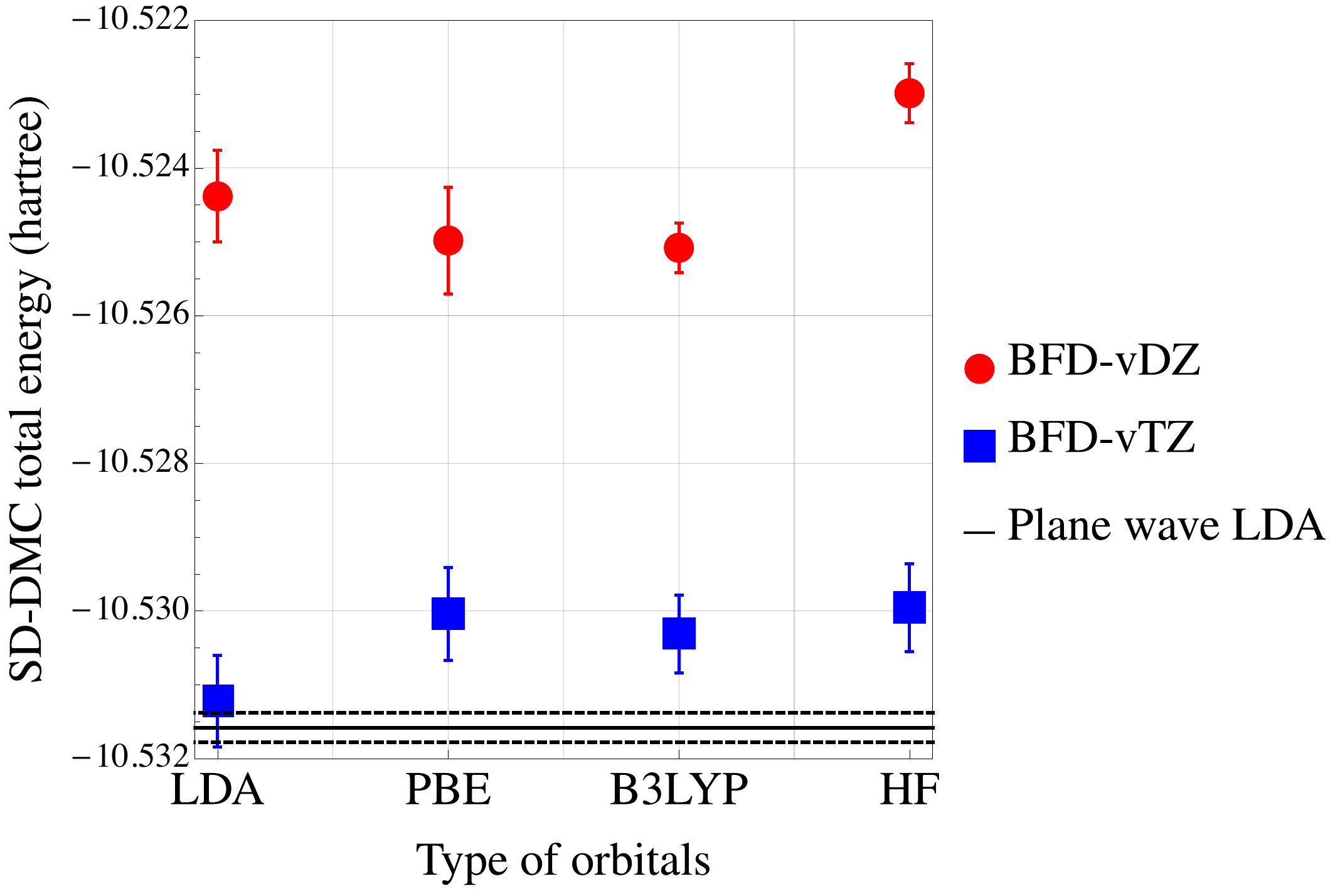}
	\caption{SD-DMC total energy of the $\cell{1}{1}{1}$ diamond primitive cell computed at the $\Gamma$ point using SD trial wavefunctions built with various orbitals (LDA, PBE, B3LYP, and HF) and the BFD-vDZ (blue) or BDF-vTZ (red) GTO basis sets. 
	The solid line corresponds to the SD-DMC energy obtained using a large PW basis set and LDA orbitals and the dashed lines indicate the associated statistical error.  
	The raw data can be found in the {\SI}.
	\label{fig:BasisSet}
	}
\end{figure}

To establish a reference for the multideterminant studies we first investigate the dependence of single determinant DMC energies on the
initial orbitals and their basis sets. In Fig.~\ref{fig:BasisSet}, we show the dependence of the fixed-node DMC energy for single
determinant trial wavefunctions built with {local-density approximation} (LDA)\cite{Perdew_1981}, {Perdew–Burke-Ernzerhof} (PBE)\cite{Perdew_1996}, {Becke, 3-parameter, Lee–Yang–Parr} (B3LYP)\cite{Becke_1993,Lee_1988,Vosko_1980,Stephens_1994}
and HF orbitals calculated with the BFD-vDZ and BFD-vTZ basis sets. These calculations were performed for the simplest case of a
single primitive cell consisting of two carbon atoms ($\cell{1}{1}{1}$ supercell), with the SD-DMC energy being computed at the
$\Gamma$ point. As an additional reference, the SD-DMC energy obtained using LDA orbitals expanded in a large plane-wave (PW)
basis set with a high energy cutoff of 175 Hartree and the same pseudopotential is also reported. Using a larger energy cutoff of 250 Hartree leads to a
difference in DMC energies below 1 milliHartree and therefore our initial choice of 175 Hartree is close to the complete basis set limit. Note
that the fixed node SD-DMC energies using nodal surfaces of the Kohn-Sham determinants obtained from different DFT approximations differ by less than the variation in the
pure DFT energies, 6.8(6) milliHartree vs 13.05 milliHartree, respectively. The SD-DMC energies obtained with LDA orbitals
expanded in the BFD-vTZ and PW basis sets coincide within {1.0(6)} milliHartree thus indicating the suitability of the BFD-vTZ basis set.  On
the other hand, an appreciably higher SD-DMC energy is obtained when using the BFD-vDZ basis set. Consequently, in the following,
all calculations are performed with the BFD-vTZ basis set.

\begin{table}
	\caption{Twist-averaged SD-DMC total energies (in Hartree/cell) and cohesive energies (in Hartree) of diamond for supercells of increasing size $N = n^3$ ($n=2$, $3$, and $4$).
	The (extrapolated) thermodynamic values and the cohesive energies (with and without ZPE correction) are also reported. 
	\label{tab:SD-total}
	}
		\begin{ruledtabular}
		\begin{tabular}{lrr}
			Cell size								&  \tabc{SD-DMC} 				& \tabc{SD-DMC}	\\ 
			($\cell{n}{n}{n}$)						&  \tabc{(B3LYP/GTO)\fnm[1]} 	& \tabc{(PBE/PW)\fnm[2]}	\\ 
			\hline
			$\cell{2}{2}{2}$	 					&	$-11.4217(1)$	&	$-11.4199(1)$		\\
			$\cell{3}{3}{3}$	 					&	$-11.4078(7)$	&	$-11.4049(1)$		\\
			$\cell{4}{4}{4}$						&	$-11.4020(5)$	&	$-11.3995(1)$	\\
			\hline
			Extrapolated ($N \to \infty$)\fnm[3] 	&	$-11.3971(7)$	&	$-11.3949(1)$	\\ 
			Cohesive energy (w/o ZPE) 				&	$0.2772(4)$	&	$0.2755(1)$ 	\\
			Cohesive energy (w/ ZPE)		&	$0.2712(4)$	&	$0.2695(1)$ 	\\
		\end{tabular}
		\end{ruledtabular}
		\fnt[1]{Results from the present work obtained with B3LYP orbitals and the BFD-vTZ GTO basis set.}
		\fnt[2]{Results from Ref.~\onlinecite{Shin_2014} obtained with PBE orbitals and a PW basis set. }
		\fnt[3]{Extrapolated values obtained using a using {a simple quadratic fit of the energies for the three sizes as a function of 1/N.}}
\end{table}

In Table \ref{tab:SD-total}, we report twist-averaged SD-DMC total energies (per primitive cell) as a function of the supercell
size $N$, as well as the corresponding cohesive energies. We also report the SD-DMC energies computed by Shin \textit{et al.}~in
Ref.~\onlinecite{Shin_2014}, which were obtained using a PW basis set and a single Slater determinant wavefunction of PBE
orbitals and using the same twist values. \footnote{Data of Ref.~\onlinecite{Shin_2014} kindly provided by the authors.} Results are compared to our SD-DMC data
using B3LYP orbitals and the BFD-vTZ basis set. The SD-DMC total energies from the two calculations are very
close, with a difference never greater than $3$ milliHartree. Electronic cohesive energies calculated by subtracting from the
crystal energy the SD-DMC energy of the isolated atoms calculated using the same approaches [$-5.4226(2)$ Hartree/atom for
PBE/PW and $-5.42138(2)$ Hartree/atom for B3LYP/GTO] lead to a difference of only $3.0\pm0.4$ milliHartree once extrapolated to
the thermodynamic limit. Adding the zero-point vibrational energy (ZPE) correction estimated to be $6.0$ milliHartree/atom
\cite{Schimka_2011} brings the SD-DMC(B3LYP/GTO) cohesive energy to $1.3\pm0.4$ milliHartree of the estimated experimental value
\cite{Chase_1985} of $0.2699$ Hartree (see Table \ref{tab:SD-total}). The result indicates that there is very good error cancelation
between the bulk and atomic total energies at the single determinant level in carbon diamond.

\subsection{Multideterminant trial wavefunctions}

\begin{figure}
	\includegraphics[width=\linewidth]{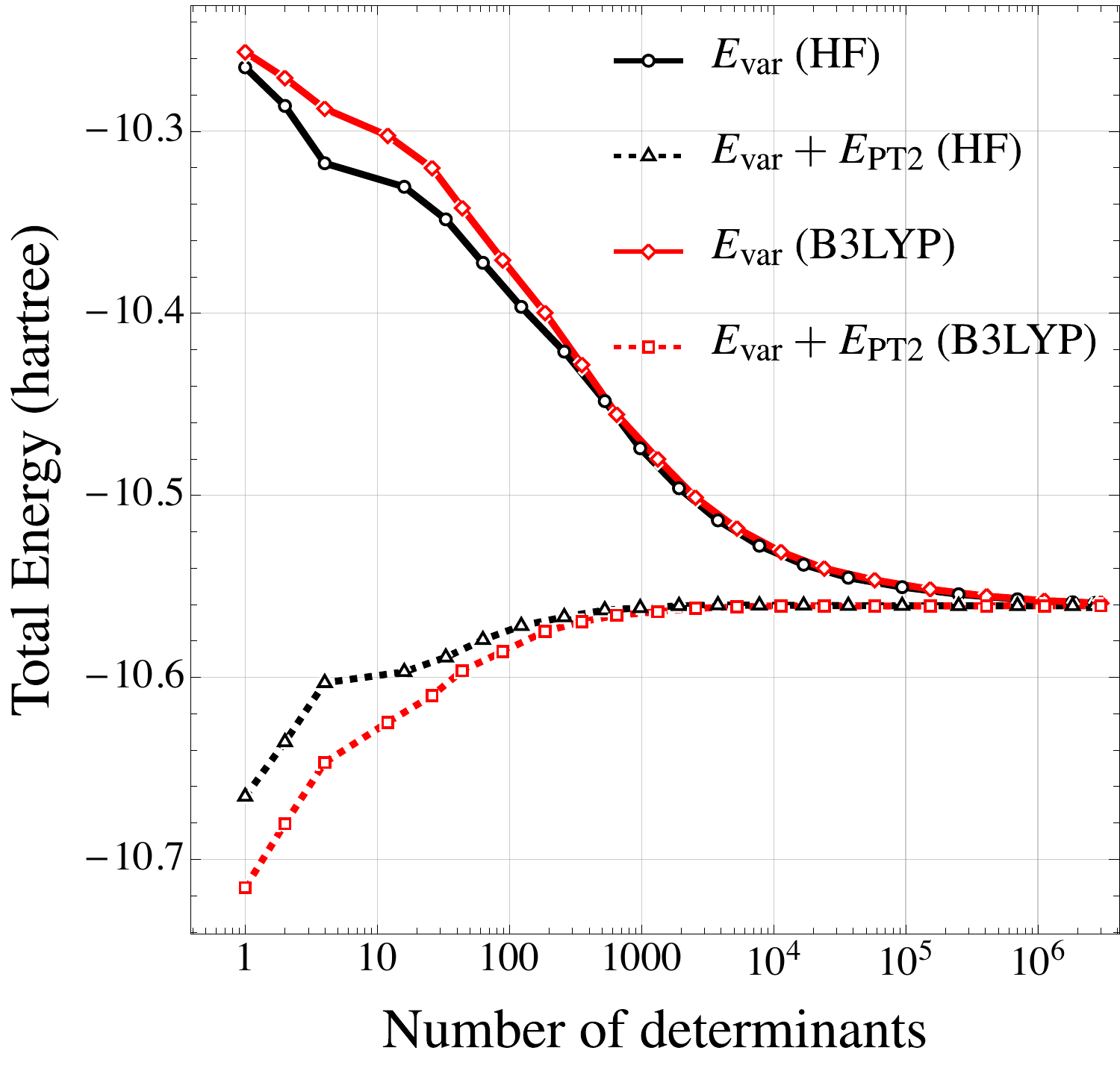}
	\caption{Convergence of the CIPSI energies with the size of the variational space for the $1\times1\times1$ diamond primitive cell computed at the $\Gamma$ point.
	Both the CIPSI variational energy $E_{var}$ and its second-order corrected value $E_{var} + E_{\rm PT2}$ (dashed) computed with the BFD-vTZ basis set using HF orbitals (black) and B3LYP orbitals (red) are reported.
	The raw data can be found in the {\SI}.
	\label{fig:Figures/ConvergingEPT2}}
\end{figure}

In Fig.~\ref{fig:Figures/ConvergingEPT2}, we show the convergence of the CIPSI variational energy $\Evar$ and its second-order
corrected value $\Evar + \EPT$ as a function of the number of determinants in the reference space (see Sec.~\ref{sec:CIPSI}) for
the $\cell{1}{1}{1}$ diamond primitive cell at the $\Gamma$ point and using the BFD-vTZ basis set. Independent calculations using
both HF and B3LYP orbitals are shown. Despite the large size of the Hilbert space (about $\sim 10^{11}$ determinants), energy
convergence is achieved with a reasonable accuracy using less than $\sim 10^{7}$determinants. In addition, the FCI limit is
independent of the orbital set employed in the SCI calculation, as it should be. In the variational calculations, convergence to
milliHartree accuracy is achieved at about $3\times 10^6$ determinants whether using HF or B3LYP orbitals.  When the perturbative
corrections to the energy are included, milliHartree accuracy is reached with only  $\sim 2\times 10^3$ and $\sim 4\times 10^3$
determinants determinants for HF and B3LYP orbitals, respectively. 

Although CIPSI rapidly converges to the FCI limit for the $1\times1\times1$ supercell ($\Gamma$ point calculations) the
exponential character of the FCI approach is still present and, for larger supercells, the number of determinants required to
reach the FCI limit to the required accuracy rapidly becomes computationally intractable. To alleviate this issue and to enable
treating larger supercells {we limit the SCI calculations to a subset of orbitals belonging to a pre-defined active space, 
thus performing in practice a CIPSI calculation for a set of $N_e$ electrons distributed among $N_o$ orbitals, denoted as CIPSI($N_e$,$N_o$), instead of a FCI one.}
Of course, there are many ways of choosing such an active space (AS). {A natural choice is to build the AS using the natural orbitals 
obtained in a preliminary CIPSI calculation. This is what has been systematically done for molecular systems in a number of works published
by some of the authors.\cite{Scemama_2018a,Scemama_2018b,Caffarel_2016a,Dash_2018,Garniron_2017b,Garniron_2018,Loos_2018b,Loos_2019,Loos_2020a,Loos_2020b}
Here, in view of the large number of orbitals needed for solids, an alternative choice could be to employ instead the natural orbitals of a preliminary 
MP2 calculation. Another possibility is to use Kohn-Sham orbitals instead of Hartree-Fock ones. Here, 
we have chosen to postpone the investigation of such important aspects to a future detailed study and we
have considered the simplest approach where only virtual orbitals with one-electron energies below a given energy threshold are included in the active space. 
Although this choice is rather crude (and certainly not optimal), it is important to emphasize that the choice of the active space for constructing 
the multideterminant trial wavefunction is expected to be less critical in DMC than when just limiting the calculation to pure CI.
High-energy orbitals are indeed expected to have a small impact on the wavefunction and, thus, on the location of the nodes.}

\begin{figure}
	\includegraphics[width=\linewidth]{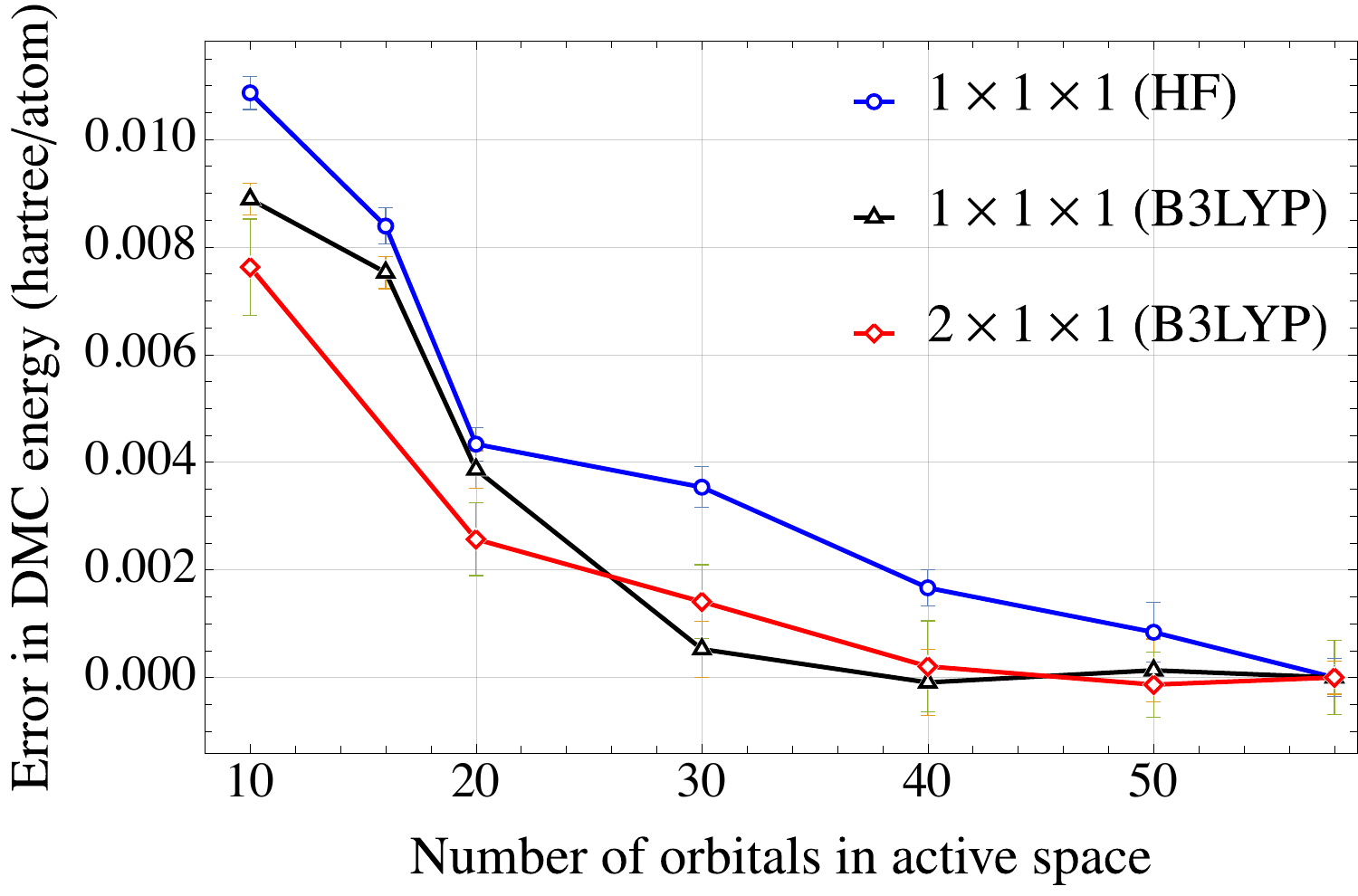}
	\caption{Error in the DMC total energy (in Hartree/atom) computed at the $\Gamma$ point as a function of 
the number of orbitals in the active space for {CIPSI($N_e$,$N_o$)} trial wavefunctions built with HF orbitals ($\cell{1}{1}{1}$ cell, 
{$N_e=8$}) or B3LYP orbitals ($\cell{1}{1}{1}$, {$N_e$=8} and $\cell{2}{1}{1}$ cells, {$N_e$=16}).
	These calculations are performed with the BFD-vTZ basis set.
	The raw data can be found in the {\SI}.
	\label{fig:DMC_CAS}}
\end{figure}

Figure \ref{fig:DMC_CAS} illustrates the use {CIPSI($N_e$,$N_o$)}. The error in DMC total energy (in Hartree/atom) computed at the
$\Gamma$ point as a function of the number of orbitals belonging to the active space is presented. The error is calculated with respect to the converged DMC value obtained with 58 orbitals. For the $\cell{1}{1}{1}$ primitive cell, results are reported
for both B3LYP and HF orbitals, and for a number of orbitals in the {CIPSI(8,$N_o$)} up to the maximum of 58. For the $\cell{2}{1}{1}$
supercell, only the B3LYP results are presented, and the maximum number of orbitals in the {CIPSI(16,$N_o$)} corresponds to only half of the
total number of orbitals ($116$). For the $\cell{1}{1}{1}$ supercell a well converged SD-DMC energy is achieved when using only 30
of the 58 B3LYP orbitals, but all orbitals must be retained to obtain a well converged result when using HF orbitals.   For the
larger $\cell{2}{1}{1}$ cell for which the {CIPSI(16,$N_o$)} is built with B3LYP orbitals, convergence to within $1.0\pm0.3$ milliHartree/atom is
reached between 30 and 40 orbitals out of a total of 116. Due to the better convergence of the calculations with truncated orbital spaces when using B3LYP than HF orbitals, B3LYP orbitals are used in all subsequent calculations. 

\subsection{Multideterminant fixed-node DMC}

\begin{figure}
	\includegraphics[width=\linewidth]{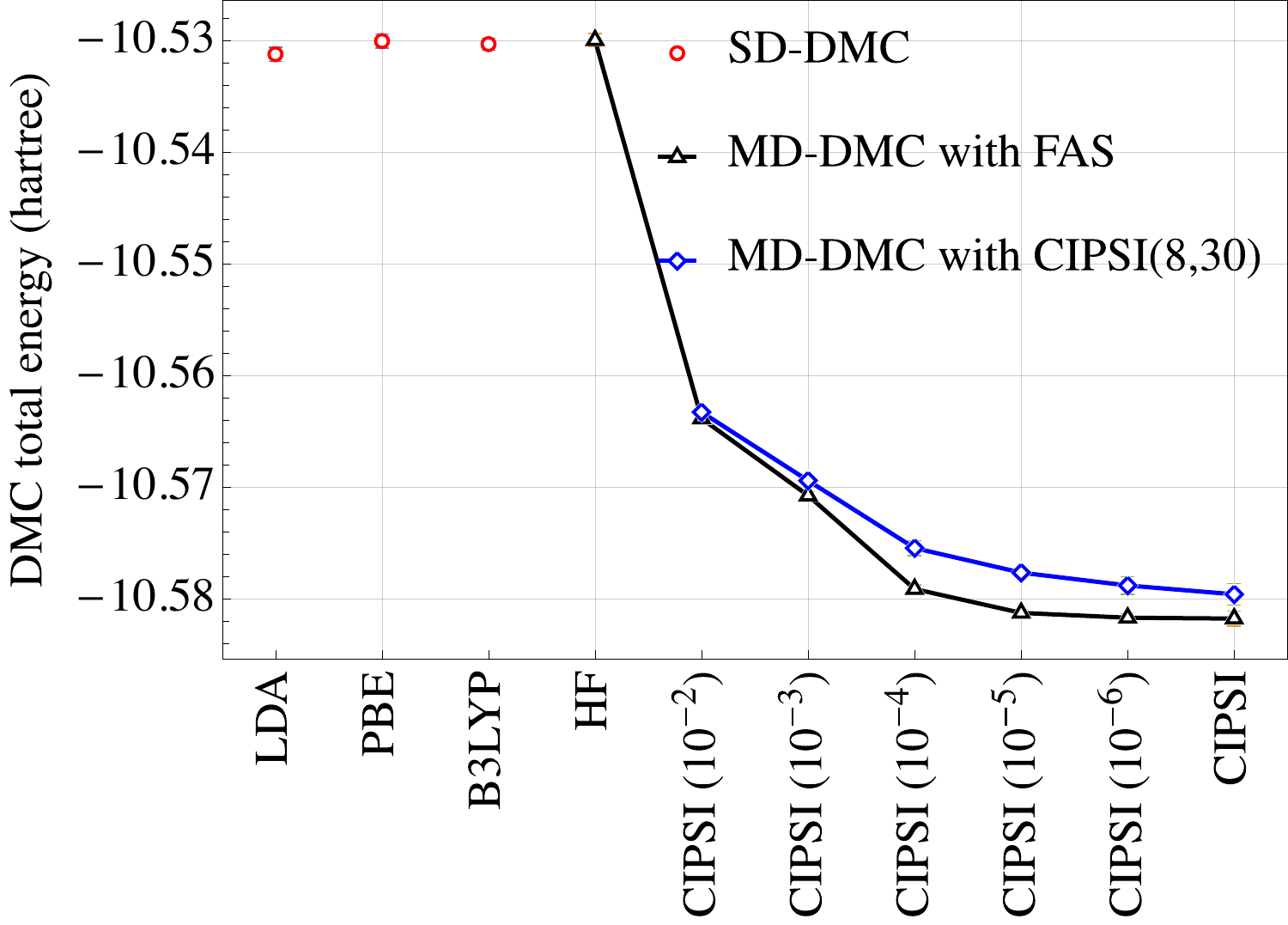}
	\caption{Comparison of DMC total energies (in Hartree) computed at the $\Gamma$ point with various SD and MD trial wave
	functions for the $\cell{1}{1}{1}$ diamond primitive cell. The SD-DMC energies (red dots) are computed with trial wave
	functions built with LDA, PBE, B3LYP, and HF orbitals. The MD-DMC energies are given as a function of the size of the CIPSI
	expansion for the FCI space which contains 58 orbitals (black curve) and the {CIPSI(8,30)} (blue
	curve). The multideterminant trial wavefunctions are built with B3LYP orbitals obtained with the BFD-vTZ basis set. Numbers in parentheses
	on the abscissa correspond to the values of the truncation threshold $\epsilon$ (see text). Error bars are either smaller or
	of the order of the size of the markers. The raw data are gathered in Table \ref{tab:Truncation} and in the {\SI}.
	\label{fig:DMC_Convergence}}
\end{figure}

Figure \ref{fig:DMC_Convergence} illustrates one of the central points of this work, namely the possibility of improving nodal
surfaces in a systematic way using SCI multideterminant trial wavefunctions. Fixed-node DMC total energies using various SD and MD trial wave
functions are represented in Fig.~\ref{fig:DMC_Convergence} for the $\cell{1}{1}{1}$ diamond primitive cell. On the left side of
the graph, DMC energies obtained with SD trial wavefunctions built with LDA, PBE, B3LYP, and HF orbitals, respectively, are
reported. (These are the same as the BFD-vTZ results reported in Fig.~\ref{fig:BasisSet}.) At the scale of the figure \ref{fig:DMC_Convergence}, the various
SD-DMC energies are very similar, and are much higher in energy than the MD-DMC energies, which are up to $0.05$ Hartree lower.
This indicates a significant improvement in the quality of the nodal surface. On the right side of the graph, the convergence of
the MD-DMC energy as a function of the size of the CIPSI expansion is presented both using the {full active space} (lower black curve) and {active space}.
(upper blue curve) spaces. The corresponding numerical data (including the number of determinants and the intrinsic variance {for DMC calculations}) are
reported in Table \ref{tab:Truncation}. Here, $\epsilon$ is a threshold associated with the number of determinants retained in the
expansion, and $\epsilon=0$ corresponds to the full CIPSI wavefunction (which is much smaller than the FCI space due to a limit on the
total number of determinants retained, see Ref.~\onlinecite{Scemama_2016}). In view of the computational cost of using a very
large number of determinants in the trial wavefunction in DMC calculations, it is important to  limit the determinants used to the
smallest number possible. As seen in Table \ref{tab:Truncation} the fixed-node energies are essentially converged within
statistical errors at $\epsilon=10^{-5}$ and, thus, to consider the full CIPSI wavefunction ($\epsilon=0$) in DMC is not necessary
here. Note that the number of determinants needed to reach a given threshold $\epsilon$ is smaller for the FCI space than the
{CIPSI(8,30)} space. This results from the greater flexibility available with the FCI space. The convergence curves of the MD-DMC
energy both for the FCI and {CIPSI(8,30)} spaces are similar and, at near-convergence, the two DMC energies differ by only about $2 \pm
1$ milliHartree. The {DMC} intrinsic variance in the energy shows a systematic improvement with the number of determinants in the
expansion when using the FCI space. 

As seen from Fig.~\ref{fig:DMC_Convergence} the DMC energy decreases monotonically and smoothly when increasing number of
determinants in the trial wavefunction. As discussed in Ref.~\onlinecite{Caffarel_2016a}, there is no guarantee that increasing
the number of Slater determinants in the trial wavefunction lowers the DMC energy, because the selection procedure
does not explicitly optimize the nodal surface and DMC energy. However in all applications performed so far -- atoms
\cite{Giner_2013,Scemama_2014,Scemama_2016}, molecules\cite{Giner_2015,Caffarel_2016a,Scemama_2018a,Scemama_2018b,Scemama_2019},
and now solids -- a systematic decrease of the fixed-node DMC energy is observed whenever the SCI trial wavefunction is improved
variationally, upon increasing the number of determinants, the size of the basis set, or the size of the active space.

\begin{table*}
\caption{DMC total energies (in Hartree/cell) computed at the $\Gamma$ point for the $\cell{1}{1}{1}$ diamond primitive cell {containing 8 electrons}, as a
	function of the truncation threshold $\epsilon$ on the coefficients of the CIPSI wavefunction computed
	within the {full active space (58 orbitals) and the active space (30 orbitals)} with the BFD-vTZ basis set. For both spaces the final CIPSI wave
	function is obtained by stopping the selection process when the number of determinants exceeds about 1 million. The number of
	determinants  in the trial wavefunction and intrinsic variance in the energy are also reported for each value of $\epsilon$
	as well as the variance of the energy inferred from the DMC data.
	\label{tab:Truncation}
	}
\begin{ruledtabular}
\begin{tabular}{lrddrdd}

	Truncation		&	\multicolumn{3}{c}{Full active space}	&	\multicolumn{3}{c}{Active space}	\\
					\cline{2-4} \cline{5-7}
	threshold $\epsilon$& \# determinants  & \tabc{DMC energy} & \tabc{Variance}& \# determinants & \tabc{DMC energy}& \tabc{Variance}	\\ 
	\hline
	$10^{-2}$		&	$178$			&	-10.5638(3)&0.2428(4)	&	$200$			&	-10.5633(2)&0.2399(5)	\\
	$10^{-3}$		&	$4124$		&	-10.5707(3)& 0.2010(1)	&	$3204$		&	-10.5694(3)&0.1950(2)	\\
	$10^{-4}$		&	$80864$		&	-10.5791(4)&0.1730(2)	&	$57990$		&	-10.5754(7)&0.1720(2) 	\\
	$10^{-5}$		&	$738998$		&	-10.5812(3)& 0.1382(7)	&	$578025$	&	-10.5776(4)& 0.16384(5)\\
	$10^{-6}$		&	$1043197$	&	-10.5817(4)&0.1372(8)	&	$1282995$	&	-10.5788(8)&0.16338(9)	\\
\hline
$\epsilon=0$	&	$1137782$	& 	-10.5817(7) &0.1363(8)	&	$1510556$	&	-10.5796(9)	&0.16260 (2)\\
\end{tabular}
\end{ruledtabular}
\end{table*}

\begin{figure}
	\includegraphics[width=\linewidth]{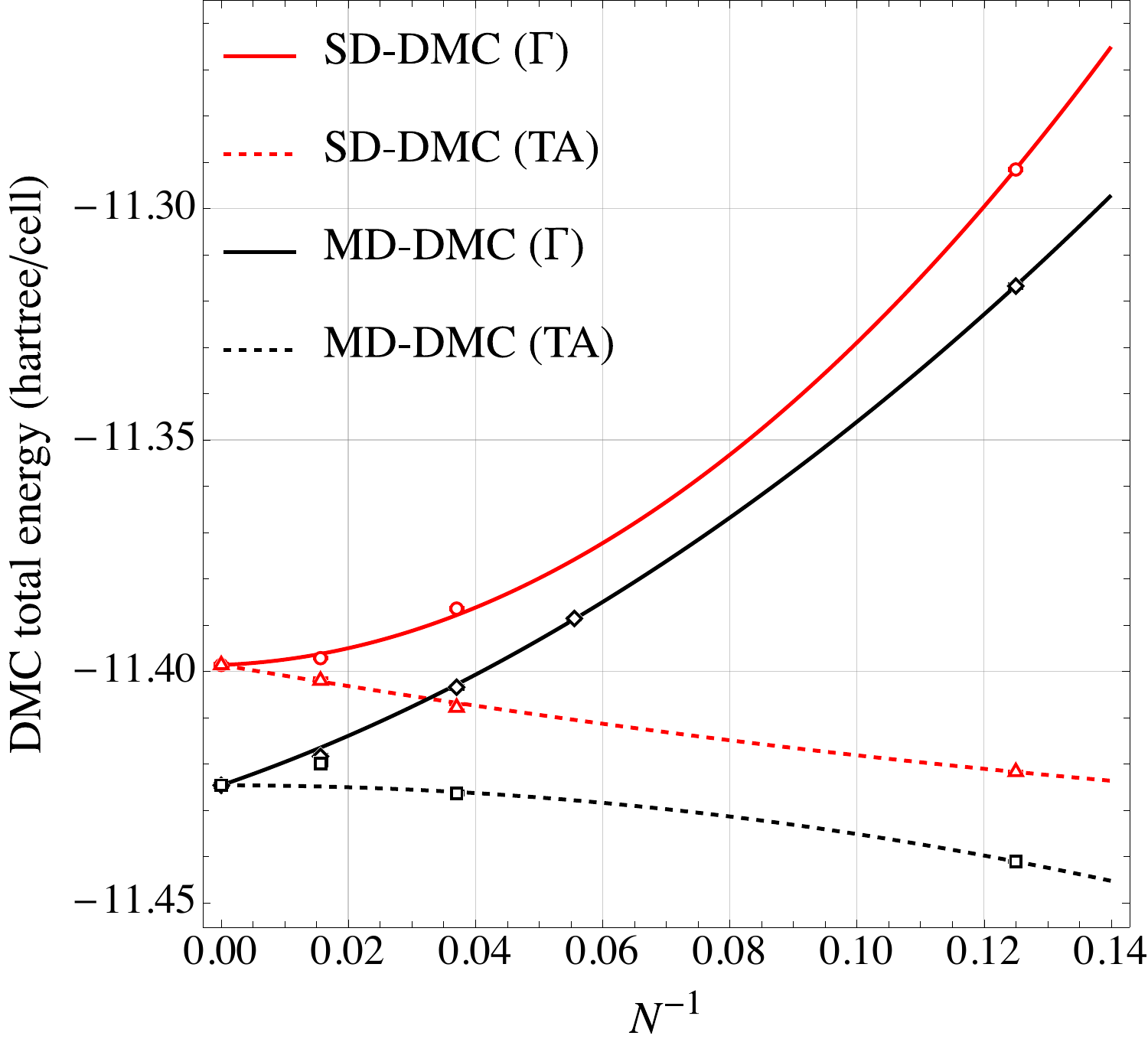}
	\caption{SD- and MD-DMC total energies (in Hartree/cell) as a function of the inverse of the system size $N$. 
	Energies are calculated either at the $\Gamma$ point ($\Gamma$, solid lines) or by twist-averaging (TA, dashed lines).
	Error bars are either smaller or of the order of the size of the markers.
	Each set of energies [SD (red lines) or MD (black lines)] are fitted simultaneously using Eq.~\eqref{eq:fit_G} or Eq.~\eqref{eq:fit_TA}.
	The raw data can be found in the {\SI}.
	\label{fig:final}
     }
\end{figure}

The DMC total energy (in Hartree/cell) of diamond as a function of $1/N$, where $N$ is the size of the system, \ie, the total
number of primitive cells replicated to create the supercell ($N = 8$, 18, 27, and 64 for $\cell{2}{2}{2}$, $\cell{3}{3}{2}$,
$\cell{3}{3}{3}$, and $\cell{4}{4}{4}$, respectively), computed with various methods and approximations is reported in
Fig.~\ref{fig:final}. The two upper curves (solid lines) report SD-DMC (red) and MD-DMC (black) energies computed at the $\Gamma$
point. The two lower curves (dashed lines) are SD-DMC (red) and MD-DMC (black) energies obtained by twist-averaging as described
in Sec.~\ref{sec:PBC}. To minimize the statistical fluctuations we re-optimized the Jastrow factor at each twist. The number of determinants
in the multideterminant trial wavefunction varies from about $600 000$ to $1 200 000$. 

SD-DMC calculations (at the $\Gamma$ point and twist-averaged) have been performed for the $\cell{2}{2}{2}$, $\cell{3}{3}{3}$, and
$\cell{4}{4}{4}$ supercells. MD-DMC calculations are much more computationally
demanding than their SD-DMC analogues.
Consequently, MD-DMC calculations have been limited to the $\cell{2}{2}{2}$ and
$\cell{3}{3}{3}$ supercells, with the energy of the $\cell{4}{4}{4}$ supercell being estimated as explained below. In all cases we only computed the symmetry inequivalent twists in the Brillouin zone, using symmetries to reduce the number of expensive DMC
calculations. There are 16, 8, and 4 inequivalent twists out of 216, 64, and 27 total for the $\cell{2}{2}{2}$, $\cell{3}{3}{3}$, and
$\cell{4}{4}{4}$ supercells, respectively.

To further reduce the computational effort, an approximate version of the
twist-averaged MD-DMC energies was employed for the $\cell{3}{3}{3}$ and $\cell{4}{4}{4}$ superlattices.
In this procedure we
estimate the energy at a given twist (other than $\Gamma$) via
\begin{equation}
	\label{eq:approx_twist}
		\EDMC(w_i) \approx \EDMC(w_0) + \EHFKS (w_i) - \EHFKS(w_0),
\end{equation}
where $w_i$ denotes the $i$th twist ($w_{0}$ being the $\Gamma$ point) and $\EHFKS$ is the energy computed with the HF energy
functional and B3LYP orbitals. This approximation relies on the B3LYP band structure being sufficiently accurate, but can be
further improved by considering how this data is extrapolated to the thermodynamic limit, as we will detail below.

\begin{figure}
	\includegraphics[width=\linewidth]{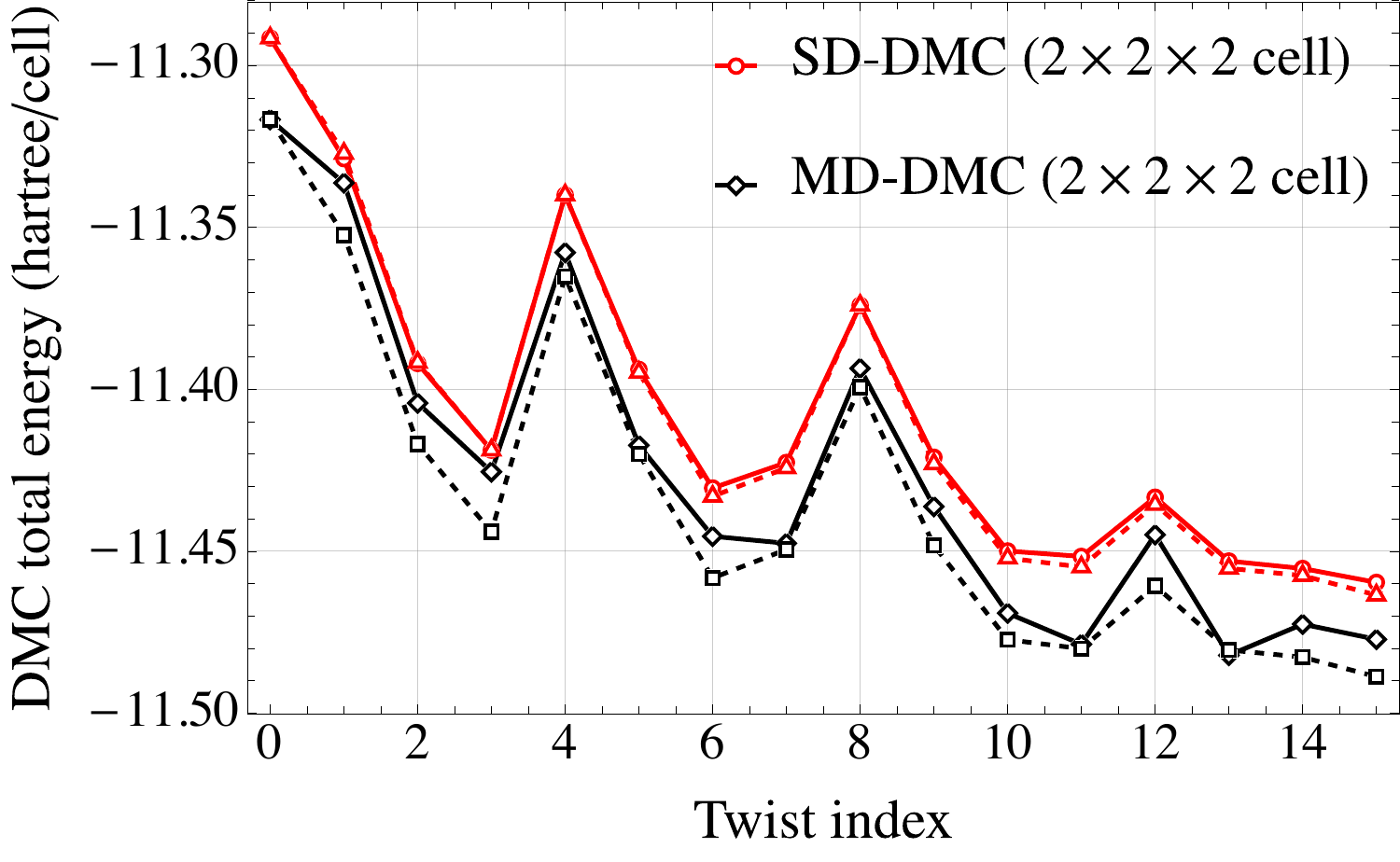}
	\caption{Twist-averaged SD-DMC (red) and MD-DMC (black) total energies (in Hartree/cell) for the $\cell{2}{2}{2}$ supercell as a function of twist index (16 independent twists out of the 216 reciprocal grid). 
	Solid and dashed lines correspond to exact and approximate calculations [using Eq.~\eqref{eq:approx_twist}], respectively.
	The DMC energies at the various twist indices are connected by straight lines only to guide the eyes.
	The raw data can be found in the {\SI}.
	\label{fig:Twist}
	}
\end{figure}

Fig.~\ref{fig:Twist} reports the SD- and MD-DMC energies computed both exactly and using Eq.~\eqref{eq:approx_twist} for the
16 independent twists of the $\cell{2}{2}{2}$ supercell. To facilitate the visualization of the actual differences between data,
the 16 DMC energies are artificially connected with straight lines. The two upper (red) curves and the two lower (black) curves
correspond to SD-DMC and MD-DMC calculations, respectively. As one can see, there is excellent agreement between the exact (solid
lines) and approximate (dashed lines) treatment of twist averaging in the case of SD-DMC{, introducing an error of only 1.7(2)mHa in the twist averaged energy}. For MD-DMC, the agreement is not as good
but remains satisfactory for the present purpose{, introducing an error of 8(4)mHa to the twist averaged energy}.

We can estimate the twist-averaged DMC energy of the $\cell{4}{4}{4}$ supercells by combining the extrapolated value of the DMC
energy at the $\Gamma$ point and a twist-averaged contribution from Eq.~\eqref{eq:approx_twist}. The extrapolated value reported
in Fig.~\ref{fig:final} has been obtained as the value resulting from a quadratic fit of the three values corresponding to the
$\Gamma$ point energies of the $\cell{2}{2}{2}$, $\cell{3}{3}{2}$, and $\cell{3}{3}{3}$ supercells. 
Note that, as is common in solid state calculations, a more precise fit function based on theoretical basis could be employed 
(see, for example ref.\onlinecite{Kolorenvc_2011,Drummond2008}). However, here in view of the remaining errors that are certainly much larger than the 
difference in error between various extrapolation fit functions, we have used the simplest possible fit function that reproduces the approximate parabolic 
shape of the curves.{
Fitting our data only within the domain of size where a quasi-linear regime is reached would clearly be preferable but the 
high computational cost of the MD-DMC data restricts the range of accessible sizes.}

{Finally, in the spirit of extracting the maximum of information from our limited set of data,}
we propose to exploit the fact that both $\Gamma$ point and twist-averaged energies must converge to the same value in the
thermodynamic limit (\ie, $N \to \infty$). This {\it exact} property is used as a constraint when fitting \textit{simultaneously} both
sets of energies with quadratic expressions, \ie, 
\begin{subequations}
\begin{align}
	\label{eq:fit_G}
	\Efit^\Gamma(N) & = \EDMC^{N \to \infty} + \frac{c_1^\Gamma}{N} + \frac{c_2^\Gamma}{N^2},
	\\
	\label{eq:fit_TA}
	\Efit^\text{TA}(N) & = \EDMC^{N \to \infty}+ \frac{c_1^\text{TA}}{N} +  \frac{c_2^\text{TA}}{N^2}.
\end{align}
\end{subequations}
The 5 parameters in Eqs.~\eqref{eq:fit_G} and \eqref{eq:fit_TA}, \ie, $\EDMC^{N \to \infty}$ and the four $c_i$'s, are obtained by
minimizing the $\chi^2$-type function 
\begin{equation}
	\chi^2 
	= \sum_i \qty[ \frac{\EDMC^\Gamma(N_i)-\Efit^\Gamma(N_i)}{\delta\EDMC^\Gamma(N_i)} ]^2
	+ 
	\sum_i \qty[ \frac{\EDMC^\text{TA}(N_i)-\Efit^\text{TA}(N_i)}{\delta\EDMC^\text{TA}(N_i)} ]^2,
\end{equation}
where the $\delta \EDMC$'s are the corresponding statistical errors, and the sum runs over $N_i$ = 8, 18, 27, and 64 for the
$\Gamma$ point energies (centered grid), and $N_i$ = 8, 27, and 64 for the twist-averaged (TA) energies. The quantity $\EDMC^{N
\to \infty}$ represents the best estimate of the DMC energy in the thermodynamic limit. 

Following this extrapolation procedure, the SD-DMC total energy in the thermodynamic limit is found to be $\ESDDMC^{N \to \infty}
= -11.3986(2)$ Hartree. Combined with the atomic SD-DMC total energy of $-5.4214(2)$ Hartree, it yields a cohesive energy
(including the ZPE contribution) of $0.2719(3)$ Hartree, an estimate close to the value of $0.2712(4)$ Hartree obtained by a
simple quadratic fit of the DMC energies computed at the $\Gamma$ point (see Table \ref{tab:SD-total}). In the multideterminant case, we find
$\EMDDMC^{N \to \infty} = -11.4246(4)$ Hartree, and using the atomic MD-DMC energy computed at $-5.4335(1)$ Hartree obtain a
ZPE-corrected cohesive energy of $0.2729(1)$ Hartree . This value compares quite well with the estimated experimental cohesive
energy of $0.2699$ Hartree extracted from Ref.~\cite{Chase_1985}. While this value is also similar to that obtained with single-determinant wavefunctions, there has been a significant lowering in the supercell and atomic energies due to use of multideterminant trial wave
functions. In particular the energy of the atom obtained from the MD-DMC calculations is essentially exact (for the chosen
pseudopotential).

It should be emphasized that comparing calculated and experimental cohesive energies is subject to the error made in estimating
the ZPE contribution. The value of $6.0$ milliHartree/atom employed in this work was obtained by Schimka \textit{et al.} using a zero-point
anharmonic expansion based on DFT energy calculations\cite{Schimka_2011}. For internal consistency and better accuracy, evaluating
the ZPE correction using the very same DMC methodology would be desirable. {Some caution is then needed with respect to the particularly good agreement between our final cohesive energies and the experimental value found in this work. Some fortuitous cancellation of systematic errors between ZPE, remaining fixed-node, and extrapolation errors could be at work.} A number of earlier {\it ab initio} calculations of the
diamond cohesive energy have been reported in the literature. Values obtained using DMC with a single-determinant
Slater-Jastrow trial wavefunction [0.2699(2) Hartree from Hood \textit{et al.}\cite{Hood_2003} and 0.2702(4) Hartree from Shin et
al.\cite{Shin_2014}] agree closely with the estimated experimental value of 0.2699 Hartree. The CCSD calculations of {McClain} et
al.\cite{McClain_2017} and Booth \textit{et al.}\cite{Booth_2013} give significantly smaller values of the cohesive energy, namely 0.2527 Hartree 
and 0.2621 Hartree, respectively. Finally, at the CCSD(T) level, 0.2712 Hartree was reported by Booth {\it et al.}\cite{Booth_2013}.
In our calculations, because the cohesive energy of the solid is slightly too large and the pseudoatom is solved essentially exactly,
the remaining difference with the experimental data must lie with a combination of the uncertainties in the extrapolation, timestep error in the DMC calculations, or pseudopotential construction and evaluation. The ZPE error is also sizable compared to the residual
difference from experiment in all of the literature predictions.

For the $\cell{2}{2}{2}$ supercell containing 16 carbon atoms, it is also instructive to compare our results to those of L{\'{o}}pez
R{\'{\i}}os \textit{et al.}\cite{Rios_2006} using the nodes of an optimized backflow (BF) trial wavefunction.
The pseudopotentials used are different, making a direct comparison of the energies not possible, although they are both around
-11.4 Hartree per primitive cell (see Table X of Ref.\onlinecite{Rios_2006} and data in our {\SI}). The BF wavefunction
improves the DMC energy by approximately 0.007 Hartree per primitive cell over the 
single-determinant result. A  truncated expansion
of 65 determinants in our $\Gamma$ point CIPSI DMC yields a reduction of 0.011 Hartree per primitive cell. A larger expansion of e.g.
about $10^6$ determinants, achieves a 0.0253(5) Hartree improvement over the SD result per primitive cell.

\section{Conclusions}\label{sec:conclusion}
We have demonstrated the feasibility of fixed-node DMC calculations for periodic solids using large multideterminant
trial wavefunctions built from SCI expansions. Using as an example carbon diamond, this procedure is shown to be able to improve
\textit{systematically} the nodal surface of the trial wavefunction. In particular, we have found that the fixed-node DMC energy
decreases monotonically and smoothly as a function of the number of determinants in the trial wavefunction.

Performing DMC calculations using large CI expansions together with large supercells is not feasible without further
approximations due to overall cost. Here, this issue was addressed by introducing an approximate, yet well-defined, protocol combining both exact and
approximate results for finite supercells and a controlled fitting procedure to reach the thermodynamic limit. Although
approximate, our estimate of the diamond cohesive energy is, to the best of our knowledge, the first example of a fully \textit{ab
initio} MD-DMC calculation of a periodic solid. In our protocol, only the Jastrow parameters are optimized in the VMC step, with
the linear coefficients of the CIPSI expansions being kept fixed. The main motivation for not optimizing the coefficients is to
exploit the property of SCI methods to provide in a simple, well-defined (linear optimization), and systematic way of
generating sequences of wavefunctions of increasing quality leading to a systematic reduction of the fixed-node error (see,
Figs. \ref{fig:DMC_CAS} and \ref{fig:DMC_Convergence}). Defining such sequences is important, for example when extrapolating
fixed-node energies at different supercell sizes or computing an energy difference such as the cohesive energy. However, the
computational cost of using (very) large CI expansions in DMC calculations is high and strategies for generating more compact
expansions are certainly desirable. Based on multideterminant DMC studies for isolated molecules (see, for example, Refs.
\onlinecite{Toulouse_2008},\onlinecite{Giner_2016b}, and \onlinecite{Dash_2019}) a practical solution consists of optimizing the CI coefficients in
presence of the Jastrow factor and keeping only the most important determinants.  However, such optimizations are challenging
because the parameters in the Jastrow factors enter non-linearly
and the objective function that one must minimize is
evaluated stochastically. From a more general perspective, it would also be desirable to investigate the comparative performance
with other types of trial wavefunctions including those with backflow as introduced by P. L{\'{o}}pez R{\'{\i}}os et
al.\cite{Rios_2006} (a first numerical comparison is presented in this work) or of geminal forms.\cite{Genovese_2020}

Finally, we note that, to exploit the full potential of the present
approach, more challenging materials must be investigated. In addition, it would be instructive to compute physical
properties other than the cohesive energy that would potentially display a different dependence on the nodal error.
We hope to be able to address these points in future studies.

\section*{Supplementary Material}\label{sec:SI}
See the {\SI} for the raw data associated with each figure of the manuscript.

\section{Data availability statement}
The data that supports the findings of this study are available within the article and its supplementary material.

\begin{acknowledgments}
Authors are grateful to Drs. Peter Doak and Qiming Sun for debugging and code implementations enabling simulations in respectively QMCPACK and PySCF.  
AS, PFL, and MC were supported by the ANR PhemSpec project, grant ANR-18-CE30-0025-02 of the French Agence Nationale de la
Recherche and the international exchange program CNRS-PICS France-USA. AB, YL, CB, JK, PK and LS were supported by the U.S.
Department of Energy, Office of Science, Basic Energy Sciences, Materials Sciences and Engineering Division, as part of the
Computational Materials Sciences Program and Center for Predictive Simulation of Functional Materials. KG and KDJ acknowledge
support from the National Science Foundation under grant CHE1762337. An award of computer time was provided by the Innovative and
Novel Computational Impact on Theory and Experiment (INCITE) program. All QMC results used resources of the Argonne Leadership
Computing Facility, which is a DOE Office of Science User Facility supported under Contract DE-AC02-06CH11357 and resources of the Oak Ridge Leadership Computing Facility, which is a DOE Office of
Science User Facility supported under Contract DE-AC05-00OR22725. For the generation
of the trial wavefunction, we gratefully acknowledge the computing resources provided on Bebop, a high-performance computing
cluster operated by the Laboratory Computing Resource Center at Argonne National Laboratory. Sandia National Laboratories is a
multi-mission laboratory managed and operated by National Technology \& Engineering Solutions of Sandia, LLC, a wholly owned
subsidiary of Honeywell International Inc., for the U.S. Department of Energy National Nuclear Security Administration under
contract DE-NA0003525.
\end{acknowledgments}

\bibliography{main}

\end{document}


\title{Supplementary Material for ``Towards a Systematic Improvement of the Fixed-Node Approximation in Diffusion Monte Carlo for Solids{ - A case study in Diamond}''}

\author{Anouar Benali}
\email{benali@anl.gov}
\affiliation{Computational Sciences Division, Argonne National Laboratory, Argonne, IL 60439, United States}
\author{Kevin Gasperich}
\affiliation{Department of Chemistry, University of Pittsburgh, Pittsburgh, Pennsylvania 15260, United States}
\author{Thomas Applencourt}
\affiliation{Argonne Leadership Computing Facility, Argonne National Laboratory, Argonne, IL 60439, United States}
\author{Ye Luo}
\affiliation{Computational Sciences Division, Argonne National Laboratory, Argonne, IL 60439, United States}
\author{M.~Chandler Bennett}
\affiliation{Materials Science and Technology Division, Oak Ridge National Laboratory, Oak Ridge, TN 37831, United States}
\author{Luke Shulenburger}
\affiliation{HEDP Theory Department, Sandia National Laboratories, Albuquerque, New Mexico 87185, USA}
\author{Paul R.~C.~Kent}
\affiliation{Center for Nanophase Materials Sciences, Oak Ridge National Laboratory, Oak Ridge, TN 37831, United States}
\affiliation{Computational Sciences and Engineering Division, Oak Ridge National Laboratory, Oak Ridge, TN 37831, United States}
\author{Jaron T.~Krogel}
\affiliation{Materials Science and Technology Division, Oak Ridge National Laboratory, Oak Ridge, TN 37831, United States}
\author{Kenneth D.~Jordan}
\affiliation{Department of Chemistry, University of Pittsburgh, Pittsburgh, Pennsylvania 15260, United States}
\author{Pierre-Fran\c{c}ois Loos}
\affiliation{Laboratoire de Chimie et Physique Quantiques, Universit\'e de Toulouse, CNRS, UPS, France}
\author{Anthony Scemama}
\affiliation{Laboratoire de Chimie et Physique Quantiques, Universit\'e de Toulouse, CNRS, UPS, France}
\author{Michel Caffarel}
\email{caffarel@irsamc.ups-tlse.fr}
\affiliation{Laboratoire de Chimie et Physique Quantiques, Universit\'e de Toulouse, CNRS, UPS, France}

\begin{abstract}
\end{abstract}

\maketitle

\begin{table}
\caption{Raw data associated with Fig.~1 of the main text.
SD-DMC energy of the $\cell{1}{1}{1}$ diamond cell in hartree computed at the $\Gamma$ point using SD trial wave functions built with various orbitals (LDA, PBE, B3LYP, and HF) and the BFD-vDZ or BDF-vTZ basis sets. 
	The SD-DMC energy obtained using a large plane wave basis set and LDA orbitals is also reported.}
\begin{tabular}{lddd}
	\hline \hline
		DMC nodal surface	&	\tabc{BFD-vDZ}		&	\tabc{BFD-vTZ}		&	\tabc{Plane wave}	\\
	\hline 
		LDA			&	-10.524\,4(6)	&	-10.531\,2(6)	&	-10.531\,6(2)	\\
		PBE			&	-10.525\,0(7)	&	-10.530\,0(6)	&   \\
		B3LYP		&	-10.525\,1(3)	&	-10.530\,3(6)	&   \\
		HF			&	-10.523\,0(4)	&	-10.530\,0(6)	\\
	\hline \hline
\end{tabular}
\end{table}

\begin{table}
\caption{Raw data associated with Fig.~2 of the main text.
	CIPSI calculations for the $\cell{1}{1}{1}$ diamond cell computed at the $\Gamma$ point.
	Zeroth-order energy $\Evar$ and its second-order corrected value $\Evar + \EPT$ computed with the BFD-vTZ basis set using HF orbitals (black) and B3LYP orbitals as a function of the number of determinants in the reference space $\Ndet$.
}
\begin{ruledtabular}
\begin{tabular}{rddrdd}
		\multicolumn{3}{c}{Trial wave function built with B3LYP orbitals}					&	\multicolumn{3}{c}{Trial wave function built with HF orbitals}					\\
		\cline{1-3} \cline{4-6}
		$\Ndet$	&	\tabc{$\Evar$}	&	\tabc{$\Evar + \EPT$}		&	$\Ndet$	&	\tabc{$\Evar$}	&	\tabc{$\Evar + \EPT$}		\\
		\hline
             1 & -10.256\,500  & -10.715\,659(0)    &          1 & -10.264\,922 &   -10.665\,580(0)		\\
             2 & -10.270\,720  & -10.680\,510(0)    &          2 & -10.286\,108 &   -10.635\,618(0)		\\
             4 & -10.287\,604  & -10.646\,809(0)    &          4 & -10.317\,600 &   -10.603\,077(0)		\\
            12 & -10.302\,464  & -10.624\,797(0)    &         16 & -10.330\,562 &   -10.597\,064(40)		\\
            26 & -10.320\,253  & -10.610\,121(68)   &         33 & -10.348\,435 &   -10.589\,059(35)		\\
            44 & -10.342\,205  & -10.596\,419(44)   &         63 & -10.372\,320 &   -10.579\,472(21)		\\
            89 & -10.370\,832  & -10.585\,930(41)   &        123 & -10.396\,530 &   -10.571\,746(25)		\\
           186 & -10.399\,763  & -10.574\,829(28)   &        260 & -10.421\,103 &   -10.566\,893(41)		\\
           354 & -10.428\,321  & -10.569\,508(30)   &        527 & -10.448\,298 &   -10.563\,121(29)		\\
           649 & -10.455\,603  & -10.565\,942(33)   &        977 & -10.474\,204 &   -10.561\,915(2)		\\
          1\,328 & -10.480\,127  & -10.564\,005(9)    &       1\,922 & -10.496\,143 &   -10.560\,772(7)		\\
          2\,564 & -10.501\,270  & -10.562\,202(5)    &       3\,786 & -10.513\,817 &   -10.560\,327(7)		\\
          5\,270 & -10.518\,185  & -10.561\,202(4)    &       7\,799 & -10.527\,755 &   -10.560\,453(3)		\\
         11\,380 & -10.531\,109  & -10.560\,784(2)    &      16\,858 & -10.538\,165 &   -10.560\,518(3)		\\
         24\,157 & -10.540\,197  & -10.560\,795(5)    &      36\,731 & -10.545\,337 &   -10.560\,666(1)		\\
         58\,210 & -10.546\,579  & -10.560\,853(10)   &      93\,865 & -10.550\,501 &   -10.560\,721(1)		\\
        153\,319 & -10.551\,656  & -10.560\,836(8)    &     251\,136 & -10.554\,414 &   -10.560\,740(4)		\\
        409\,317 & -10.555\,405  & -10.560\,813(8)    &     699\,998 & -10.557\,172 &   -10.560\,746(7)		\\
       1\,120\,101 & -10.557\,898  & -10.560\,786(9)    &    1\,831\,452 & -10.558\,815 &   -10.560\,770(9)		\\
       3\,000\,000 & -10.559\,339  & -10.560\,769(6)    &    2\,684\,325 & -10.559\,261 &   -10.560\,757(10)		\\
\end{tabular}
\end{ruledtabular}
\end{table}       

\begin{table}
\caption{Raw data associated with Fig.~3 of the main text.
	Error in DMC energy (in hartree/atom) as a function of the number of orbitals active space {CIPSI($N_e$,$N_o$)} for trial wave functions built with HF orbitals ($\cell{1}{1}{1}$ cell, {$N_e$=8}) or B3LYP orbitals ($\cell{1}{1}{1}$ and $\cell{2}{1}{1}$ cells, {$N_e$=8,16}).
	All these calculations are performed with the BFD-vTZ basis set.
}
\begin{ruledtabular}
\begin{tabular}{crdrdrd}
					&	\multicolumn{2}{c}{$\cell{1}{1}{1}$ (HF orb.)}	&	\multicolumn{2}{c}{$\cell{1}{1}{1}$ (B3LYP orb.)}	&	\multicolumn{2}{c}{$\cell{2}{1}{1}$ (B3LYP orb.)}	\\
	\cline{2-3} \cline{4-5} \cline{6-7}	
	Size of {Active Space}	&	\tabc{$\Ndet$}		&	\tabc{DMC energy}		&	\tabc{$\Ndet$}		&	\tabc{DMC energy}	&	\tabc{$\Ndet$}		&	\tabc{DMC energy}	\\
	\hline 
	10  &	 5\,319     &	 -10.560\,0(4) &	   9\,296      &	   -10.562\,8(4)	&	1\,168\,596    &	-21.866\,8(15)	\\
	16  &	 188\,120   &	 -10.564\,9(5) &	 165\,343      &	   -10.565\,6(4)  \\	
	20  &	 702\,164   &	 -10.573\,0(4) &	 538\,259      &	   -10.572\,9(6)	&	1\,996\,113     &	-21.887\,1(9)	\\
	30  &	 1\,020\,290  &	 -10.574\,6(6) &	 1\,510\,556     &	   -10.579\,6(9)	&	1\,072\,503     &	-21.891\,7(10)	\\
	40  &	 1\,000\,000  &	 -10.578\,4(4) &	 1\,332\,184     &	   -10.580\,8(12)	&	2\,355\,783     &	-21.896\,5(14)	\\
	50  &	 1\,666\,328  &	 -10.580\,0(10) &	 1\,211\,711     &	   -10.580\,4(11)	&	2\,000\,000     &	-21.897\,9(7)	\\
	58  &	 1\,831\,452  &	 -10.581\,7(5) &	 1\,137\,782     &	   -10.580\,6(4)	&	1\,827\,608     &	-21.897\,4(10)	\\
\end{tabular}
\end{ruledtabular}
\end{table}

\begin{table}
\caption{Raw data associated with Fig.~5 of the main text.
	SD- and MD-DMC total energies (in hartree/cell) as a function of the system size $N$. 
	Energies calculated at the $\Gamma$ point or by twist-averaging.
	Each set of energies (SD or MD) are fitted simultaneously (see main text for more details).}
\begin{ruledtabular}
\begin{tabular}{ccdddd}

	Cell 				&	System size			&	\multicolumn{2}{c}{SD-DMC energies}		&	\multicolumn{2}{c}{MD-DMC energies}					\\
						\cline{3-4} \cline{5-6}
	($\cell{n}{n}{n}$)	&	($N = n^3$)		&	\tabc{$\Gamma$ point}	&	\tabc{Twist-averaged}	&	\tabc{$\Gamma$ point}	&	\tabc{Twist-averaged}	\\
	\hline
	$\cell{2}{2}{2}$	&	8					&	-11.291\,5(2)		&	-11.421\,73(8) 			&	-11.316\,8(4)			&		-11.441\,11(8)	\\
	$\cell{3}{3}{2}$	&	 18					&						&	 						&	-11.388\,6(2) 			&	      					\\
	$\cell{3}{3}{3}$	&	 27					&	-11.386\,4(2)		&	-11.407\,8(7)			&	-11.403\,4(3)			&		-11.426\,4(4)	\\
	$\cell{4}{4}{4}$	&	 64					&	-11.397\,2(1)		&	-11.402\,0(5) 			&	-11.418\,4(4)			&		-11.419\,9(4)  	\\
   Extrapolated 		&	$\infty$			&	-11.398\,6(2)		&	-11.398\,6(2)	 		&	-11.424\,6(4)			&		-11.424\,6(4)		\\
\end{tabular}
\end{ruledtabular}
\end{table}

\begin{table}
\caption{Raw data associated with Fig.~6 of the main text.
	Twist-averaged SD-DMC and MD-DMC energies (in hartree/cell) for the $\cell{2}{2}{2}$ supercell as a function of the twist index (16 independent twists out of the 216 $k$-point grid), their coordinates (angles) in the supercell, and their weights based on the symmetry of the cell. 
	The exact and approximate calculations computed are reported (see main text).}
\begin{ruledtabular}
\begin{tabular}{rcddddd}
		\multicolumn{3}{c}{Twist}	&	\multicolumn{2}{c}{SD-DMC energy}	&	\multicolumn{2}{c}{MD-DMC energy}		\\
			\cline{1-3}			\cline{4-5} \cline{6-7}
	 Index 	& \tabc{Angle}	&\tabc{Weight}&\tabc{Exact} 	&	\tabc{Approx.}	&	\tabc{Exact}		&	 \tabc{Approx.}	\\
        \hline
           0 & (0.00000000,  0.00000000,  -0.00000000)&1 &-11.291\,5(2)  &     -11.291\,5(2)    &        -11.316\,7(4)  &     -11.316\,8(4)      	\\ 
           1 & (0.07761248,  0.07761248,  -0.07761248)& 8&-11.328\,8(3)  &     -11.327\,2(3)    &        -11.336\,3(2)  &     -11.352\,4(2)       \\
           2 & (0.15522495,  0.15522495,  -0.15522495)&8 &-11.392\,1(5)  &     -11.391\,7(5)    &        -11.404\,3(4)  &     -11.416\,9(4)       \\
           3 & (0.23283743,  0.23283743,  -0.23283743)&4 &-11.418\,8(3)  &     -11.418\,8(3)    &        -11.425\,5(2)  &     -11.444\,0(2)       \\
           4 & (0.00000000,  0.15522495,  0.00000000)& 6&-11.340\,0(2)  &     -11.340\,0(2)    &        -11.357\,8(1)  &     -11.365\,2(1)       \\
           5 & (0.07761248,  0.23283743,  -0.07761248)& 24&-11.393\,9(3)  &     -11.394\,8(3)    &        -11.417\,3(3)  &     -11.420\,0(3)       \\
           6 & (0.15522495,  0.31044990,  -0.15522495)&24 &-11.430\,5(2)  &     -11.433\,1(2)    &        -11.445\,4(3)  &     -11.458\,3(3)       \\
           7 & (0.69851228,  0.85373723,  -0.69851228)& 24&-11.422\,6(3)  &     -11.424\,3(3)    &        -11.447\,6(3)  &     -11.449\,5(3)       \\
           8 & (0.77612476,  0.93134971,  -0.77612476)&12 &-11.374\,1(2)  &     -11.374\,1(2)    &        -11.393\,6(5)  &     -11.399\,3(5)       \\
           9 & (0.00000000,  0.31044990,  0.00000000)& 6&-11.421\,0(4)  &     -11.423\,0(4)    &        -11.436\,2(3)  &     -11.448\,2(3)       \\
          10 & (0.07761248,  0.38806238,  -0.07761248)&24 &-11.450\,0(3)  &     -11.452\,1(3)    &        -11.469\,2(2)  &     -11.477\,4(2)       \\
          11 & (0.62089981,  0.93134971,  -0.62089981)& 12&-11.451\,6(4)  &     -11.454\,9(4)    &        -11.478\,8(4)  &     -11.480\,1(4)       \\
          12 & (0.155224951,  0.31044990,  0.00000000)& 24&-11.433\,4(3)  &     -11.435\,6(3)    &        -11.444\,9(3)  &     -11.460\,8(3)       \\
          13 & (0.69851228,  0.85373723,  -0.54328733)& 24&-11.453\,1(1)  &     -11.455\,3(1)    &        -11.482\,2(2)  &     -11.480\,5(2)       \\
          14 & (0.00000000,  0.46567485,  0.000000000)& 3&-11.455\,4(2)  &     -11.457\,6(2)    &        -11.472\,6(2)  &     -11.482\,8(2)       \\
          15 & (0.62089981,  0.93134971,  -0.46567485)&12 &-11.459\,7(3)  &     -11.463\,6(3)    &        -11.477\,2(4)  &     -11.488\,8(4)       \\
\end{tabular}
\end{ruledtabular}
\end{table}

\begin{table*}
\caption{DMC total energies (in hartree/cell) computed at the $\Gamma$ point for the $\cell{1}{1}{1}$ diamond primitive cell as a function of the truncation threshold $\epsilon$ on the coefficients of the full CIPSI wave function ($\epsilon=0$) computed within the FCI space with the BFD-vTZ basis set. The full CIPSI wave function is obtained by stopping the selection process when the number of determinants exceeds about 1 million. The number of 	determinants  in the trial wave function, the intrinsic variance the number of blocks to reach DMC convergence the efficiency of the run  and the total run time are also reported for each value of $\epsilon$. we define the efficiency as $\frac{1}{\sigma ^2t}$, where $\sigma$ is the error bar and t is the total time (number of walkers $\times$ wall time).
	\label{tab:truc-full-CI-timing}
	}
\begin{ruledtabular}
\begin{tabular}{lrddrrd}

	Truncation		&	\multicolumn{6}{c}{FCI space}\\
					\cline{2-7} 
	threshold $\epsilon$& \# determinants  & \tabc{DMC energy} & \tabc{Intrinsic variance}& \# blocks &  \tabc{Efficiency} & \tabc{KNL core-hours}	\\ 
	\hline
	$10^{-2}$		&	$178$			&	-10.5638(3)&0.2428(4)	&	$949$			&	268.89  & 2457	\\
	$10^{-3}$		&	$4124$		&	-10.5707(3)& 0.2010(1)	&	$949$		&	183.29	&3550\\
	$10^{-4}$		&	$80864$		&	-10.5791(4)&0.1730(2)	&	$646$		&31.81 	&11741\\
	$10^{-5}$		&	$738998$		&	-10.5812(3)& 0.1382(7)	&	$335$	&8.80&32768\\
\hline
	$\epsilon=0$		&	$1043197$	&	-10.5817(4)&0.1372(8)	&	$194$	&9.00&	32768\\
\end{tabular}
\end{ruledtabular}
\end{table*}

\begin{table*}
\caption{DMC total energies (in hartree/cell) computed at the $\Gamma$ point for the $\cell{1}{1}{1}$ diamond primitive cell as a function of the truncation threshold $\epsilon$ on the coefficients of the full CIPSI wave function ($\epsilon=0$) computed within {CIPSI(8,30)} space with the BFD-vTZ basis set. The full CIPSI wave function is obtained by stopping the selection process when the number of determinants exceeds about 1 million. The number of 	determinants  in the trial wave function, the intrinsic variance the number of blocks to reach DMC convergence, the efficiency of the run and the total run time are also reported for each value of $\epsilon$. we define the efficiency as $\frac{1}{\sigma ^2t}$, where $\sigma$ is the error bar and t is the total time (number of walkers $\times$ wall time).
	\label{tab:truc-cas-CI-timing}
	}
\begin{ruledtabular}
\begin{tabular}{lrddrrd}

	Truncation		&	\multicolumn{6}{c}{{CIPSI(8,30)}}\\
					\cline{2-7} 
	threshold $\epsilon$& \# determinants  & \tabc{DMC energy} & \tabc{Intrinsic variance}& \# blocks & \tabc{Efficiency} & \tabc{KNL core-hours}	\\ 
	\hline

	$10^{-2}$		&		$200$		&	-10.5633(2)&0.2399(5)&949	&553.30&2457\\
	$10^{-3}$		&		$3204$		&	-10.5694(3)&0.1950(2)&	949&535.32&3003\\
	$10^{-4}$		&		$57990$		&	-10.5754(7)&0.1720(2)& 195&518.59	&17476\\
	$10^{-5}$		&		$578025$	&	-10.5776(4)& 0.16384(5)&600&7.49&32768\\
\hline
	$\epsilon=0$		&	$1282995$	&	-10.5788(8)&0.16338(9)&206&2.17	&32768\\
\end{tabular}
\end{ruledtabular}
\end{table*}
\bibliography{all}